\documentclass{article}
\usepackage[a4paper,margin=1in]{geometry}
\usepackage{graphicx} 
\graphicspath{{Figures/}}
\usepackage{authblk}
\usepackage{comment}
\usepackage{booktabs}
\usepackage{amsmath}
\usepackage{mathtools}
\usepackage{amssymb}
\usepackage{tabularx}
\newcolumntype{Y}{>{\centering\arraybackslash}X}
\usepackage{xcolor}
\usepackage{array}
\usepackage[numbers]{natbib}
\usepackage{lineno}

\usepackage{soul}
\sethlcolor{yellow}

\newif\ifshowhighlight
\showhighlightfalse   

\ifshowhighlight
  \newcommand{\rev}[1]{\hl{#1}}
\else
  \newcommand{\rev}[1]{#1}
\fi

\newcommand{\vor}{\pmb{\omega}}
\newcommand{\pres}{\pmb{p}}
\newcommand{\Pres}{\pmb{P}}
\newcommand{\lift}{C_L}
\newcommand{\lat}{\pmb{\xi}}
\newcommand{\Lat}{\pmb{\Xi}}
\newcommand{\weights}{\pmb{W}}
\newcommand{\h}[1]{\hat{#1}}
\newcommand{\force}{\pmb{F}}
\newcommand{\encoder}{\mathcal{F}_e}
\newcommand{\decoder}{\mathcal{F}_d}
\newcommand{\obs}{\pmb{h}_{\weights}}
\newcommand{\frw}{\pmb{f}_{\weights}}
\newcommand{\frww}{\tilde{\pmb{f}}_{\weights}}
\newcommand{\obsnoise}{\pmb{\eta}}
\newcommand{\frwnoise}{\pmb{w}}
\newcommand{\gain}{\pmb{K}}
\newcommand{\stateCov}{\pmb{\Sigma}_{\lat}}
\newcommand{\obsCov}{\pmb{\Sigma}_{\obsnoise}}
\newcommand{\obsCrossCov}{\pmb{\Sigma}_{\pmb{p} \pmb{p}}}
\newcommand{\stateObsCov}{\pmb{\Sigma}_{\pmb{\xi} \pmb{p}}}
\newcommand{\stateGram}{\pmb{C}_{\lat}}
\newcommand{\obsGram}{\pmb{C}_{\pres}}
\newcommand{\stateMode}{\pmb{V}}
\newcommand{\stateEig}{\pmb{\Lambda}_{\lat}}
\newcommand{\obsMode}{\pmb{U}}
\newcommand{\obsEig}{\pmb{\Lambda}_{\pres}}
\newcommand{\stateRank}{r_{\lat}}
\newcommand{\obsRank}{r_{\pres}}

\newcommand{\priorCov}{\pmb{\Sigma}_{f}}

\newcommand{\hlin}{\pmb{H}}

\title{Sequential estimation of disturbed aerodynamic flows from sparse measurements via a reduced latent space}
\author[1,*]{Hanieh Mousavi}
\author[1]{Anya Jones}
\author[1]{Jeff Eldredge}
\affil{Mechanical and Aerospace Engineering, University of California, Los Angeles, Los Angeles, CA 90095-1597, USA}
\affil[*]{Corresponding author, email: hnmousavi@ucla.edu}
\date{}

\begin{document}

\maketitle

\begin{abstract}
\rev{This work presents a fast and uncertainty-aware sequential data assimilation framework suitable for estimation of key aerodynamic states (e.g., instantaneous vorticity fields and aerodynamic loads) during severe gust encounters, where vortex–gust interactions strongly affect the flow dynamics. The framework comprises an ensemble Kalman filter (EnKF), designed to detect and reconstruct nearly-impulsive flow disturbances with a wide range of strengths and orientations and introduced at arbitrary times. The forecast and measurement update (analysis) stages of the EnKF are each composed of learned operators in a low-dimensional latent space, obtained via a physics-augmented autoencoder. The forecast operator propagates the undisturbed baseline dynamics, but cannot predict random gust-induced deviations from these dynamics. Thus, the analysis stage frequently assimilates new surface pressure measurements to listen for disturbance signals and initiate deviation from the nominal trajectory. The methodology is trained and tested on flowfield snapshots from high-fidelity simulations of two-dimensional airfoil-gust encounters and corresponding sparse surface pressure data. Because assimilation occurs entirely within the reduced-order latent space, the updates are computationally efficient and ensure that aerodynamic states can be continuously estimated from streaming pressure data. The estimated instantaneous latent state remains physically interpretable via decoding to the original high-dimensional flow. Eigenvalue decomposition of the state and observation Gramians in the measurement update stage reveals the dominant correction directions necessary to capture the flow disturbance and quantifies how sensors inform the state corrections throughout gust interaction. The framework readily accounts for sensor failure; sensor dropout experiments show that the EnKF adaptively re-weights neighboring sensors to compensate for lost information, preserving estimation quality even under degraded sensing configurations.}
\end{abstract}

\section{Introduction}\label{sec:introduction}
Aerial disturbances, particularly those arising from various gust encounters, frequently disrupt the nominal operation and controlled maneuvering of small air vehicles. In gust-passage scenarios, the disturbance acts over a limited time window, creating sharp, transient deviations in the flow field that are especially difficult to detect and predict in real time. These disruptions can be mitigated through active control strategies---such as linear feedback control or reinforcement learning (RL)---that enable the system to respond adaptively \citep{li2022review}. However, effective active control requires real-time access to the underlying states of the dynamical system to detect anomalies and take corrective action. In aerodynamic applications, these states are typically not directly measurable. Instead, transient flow variations are reflected in (and partially observable through) the surface sensors, such as pressure taps mounted on the wings. These measurements serve as indirect observables that can be sequentially assimilated to infer the hidden flow states. In such scenarios, without a good initial guess, it is impossible to infer the inherently high-dimensional flow field from a single measurement taken from a limited number of sensors. Though sequential filters, such as those as the heart of this work, leverage past measurements and dynamical predictions to progressively narrow the search region, each measurement update of the state is still heavily challenged to navigate the high-dimensional space. 

To address this challenge, a low-dimensional state space can be constructed to serve as a compressed representation of the full flow field, describing only the most energetic structures and essential physics. Traditional approaches rely on linear projection techniques, such as Proper Orthogonal Decomposition (POD) \citep{berkooz1993proper,kaveh2025data} and Dynamic Mode Decomposition (DMD) \citep{schmid2010dynamic}, which approximate the system in terms of dominant linear modes. However, these methods have been shown to be inadequate for classes of strongly nonlinear systems---such as gust-disturbed flows---where the underlying dynamics often evolve on a nonlinear manifold \citep{fukami2023grasping}. In such settings, the modes excluded by linear truncation can significantly affect the system's evolution \citep{maulik2020time}. \rev{For gust–disturbed flow over an airfoil at low Reynolds number ($Re=100$), \mbox{\citet{fukami2023grasping}} demonstrated that the POD coordinates in the learned latent space cause dynamically distinct vorticity regimes to collapse onto overlapping trajectories. In contrast, the same study demonstrated that nonlinear autoencoders are capable of constructing latent representations that preserve the dynamical distinctions between flow states, enabling accurate low-dimensional compression and reconstruction of transient gust–airfoil interactions. Motivated by this improved expressiveness, nonlinear data compression techniques based on deep learning (DL), such as standard autoencoders \mbox{\citep{fukami2023grasping,xie2024smooth,mousavi2025low}} and their variational counterparts (VAEs) \mbox{\citep{fraccaro2017disentangled,he2025physics}}, have gained traction for their ability to learn compact, task-specific latent representations that more faithfully capture the transient dynamics of complex flow systems.}

Once a reduced-order latent space is learned, data assimilation (DA) can be performed efficiently within this space to enable fast posterior state estimation with low computational overhead using sequential filtering. From the estimated latent state and its uncertainty, the corresponding estimates in the full flow field state space can be reconstructed via the decoder. Typically, filtering frameworks involve two critical components to evolve the probability distribution of states: a \emph{forecast operator} that propagates states forward in time, and an \emph{observation operator} that maps states to measurable outputs, enabling state correction based on new observations \citep{asch2016data,eldredge2025practical}. In a reduced space, these operators can be approximated by DL-based surrogate models, providing a tractable alternative to traditional high-fidelity solvers \citep{bach2024inverse}. With the abundance of data, neural networks (NNs), in particular, are well-suited for this task as they are designed to be differentiable, enabling gradient-based optimization, and they can either augment or fully replace physics-based models in scenarios where simulations are expensive or unavailable \citep{jin2021new}.

Several recent works have demonstrated the successful use of NNs for learning surrogate models. For instance, \citet{fraccaro2017disentangled} developed the Kalman Variational Autoencoder (Kalman-VAE), which uses a recurrent neural network (RNN) to model latent dynamics from partially observed image sequences. \citet{maulik2020time} employed long short-term memory (LSTM) networks to model the evolution of unresolved POD modes in turbulent flows. In a Kalman filtering setting, \citet{coskun2017long} used three separate LSTM models to jointly learn the latent dynamics and the associated process and measurement noise statistics. Other works embed physical constraints into the learned latent dynamics: \citet{popov2022multifidelity} and \citet{he2025physics} constructed physics-informed surrogate models constrained by PDE structure. The sequence prediction models employed in the aforementioned references leverage the full temporal history to forecast future states, often using RNNs with memory kernels \citep{tang2020deep, cheng2022data}.

Other studies have integrated DL techniques into traditional filtering frameworks for improved state estimation. \citet{jouaber2021nnakf} introduced a neural Kalman filter in which an LSTM dynamically learns the process noise covariance, with the full Kalman update embedded in the computational graph. \citet{song2022improved} proposed KFFLSTM, a hybrid architecture that jointly learns dynamics and noise models using LSTMs. \citet{revach2022kalmannet} developed KalmanNet, which learns the Kalman gain in extended Kalman filtering with known models but unknown noise statistics, albeit without explicit uncertainty quantification. In large-scale geophysical applications, \citet{wang2024deep} used NNs to enhance ensemble adjustment Kalman filters for strongly coupled ocean–atmosphere systems, and \citet{howard2024machine} developed a CNN-based surrogate to accelerate the ensemble Kalman filter (EnKF) analysis step in high-resolution observational settings. Besides model-based filtering, fully model-free Bayesian estimators have also been introduced: \citet{ghosh2024danse} proposed DANSE, an RNN-based Bayesian filter that infers posterior distributions from noisy observations, and they further extended it to a semi-supervised version, SemiDANSE  \citep{ghosh2024data}, incorporating state supervision as a regularization term. \rev{A recent line of work by \mbox{\citet{ozalp2025real}} and \mbox{\citet{ozan2025data1, ozan2025data2}} has advanced data assimilation and control of chaotic flows by integrating nonlinear autoencoders, latent dynamical models, filtering, and reinforcement learning, resulting in unified frameworks for forecasting and control in partially observed chaotic systems.} Beyond filtering, variational DA approaches like 3D-Var and 4D-Var remain widely used for deterministic state estimation via optimization of forecast-observation mismatches \citep{le1986variational, kalnay2003atmospheric, de2022coupled, bach2024inverse}. However, these variational methods yield only point estimates and lack online adaptability, making them unsuitable for control-driven aerodynamic applications requiring fast and probabilistic updates.
While these approaches provide powerful filtering frameworks, they have not been developed with the specific challenges of aerodynamic gust detection, including high-dimensional flow states, sparse pressure sensing, and disturbance onset occurring at unknown times.

\rev{The current work introduces a data assimilation framework for unsteady aerodynamic flows, designed to detect and reconstruct sudden gust encounters from sparse surface pressure measurements. Each encounter originates from the same undisturbed baseline flow state, and because gust disturbances are introduced at arbitrary times, a learned forecast model alone cannot anticipate the transition from the nominal baseline trajectory to a gust-driven response without a priori knowledge of the gust parameters. In the absence of measurement updates, the forecast would simply continue to propagate the undisturbed dynamics. The central contribution of this study is the systematic use of the Kalman filtering framework---specifically its analysis step---to leverage surface pressure sensors as transient disturbance detectors that identify and correct for gust encounters of varying strength and orientation occurring at arbitrary times. Through frequent measurement updates, the filter continuously assimilates sensor information, enabling rapid detection of deviations from the baseline flow and reconstruction of the full aerodynamic response.
This sequential estimation mechanism is applied to a compact latent representation of the high-dimensional vorticity field extracted by a nonlinear physics-augmented convolutional autoencoder. Both the forecast and observation operators are learned in the latent space and incorporated into the ensemble Kalman filter (EnKF), enabling computationally efficient updates while preserving a physically interpretable uncertainty structure. Further eigendecomposition of the state and measurement Gramians enables quantification of the instantaneous informativeness of individual sensors as gust vortices convect over the airfoil and interact with vortices from the wing and in the wake. The resulting methodology provides scalable, uncertainty-aware reconstruction of gust-driven aerodynamic dynamics from sparse measurements.}

The paper is organized as follows: Section~\ref{sec:problem} formulates the problem and outlines the mathematical framework that underpins our study. Section~\ref{sec:results} reports results on complex, transient, disturbance-driven aerodynamic cases, demonstrating efficient online state estimation with the approach. Finally, Section~\ref{sec:conclusion} summarizes the main findings and discusses limitations, directions, and suggestions for future work.

\section{Problem description and methodology}\label{sec:problem}
\begin{figure*}
\centering
\includegraphics[width=1.0\textwidth]{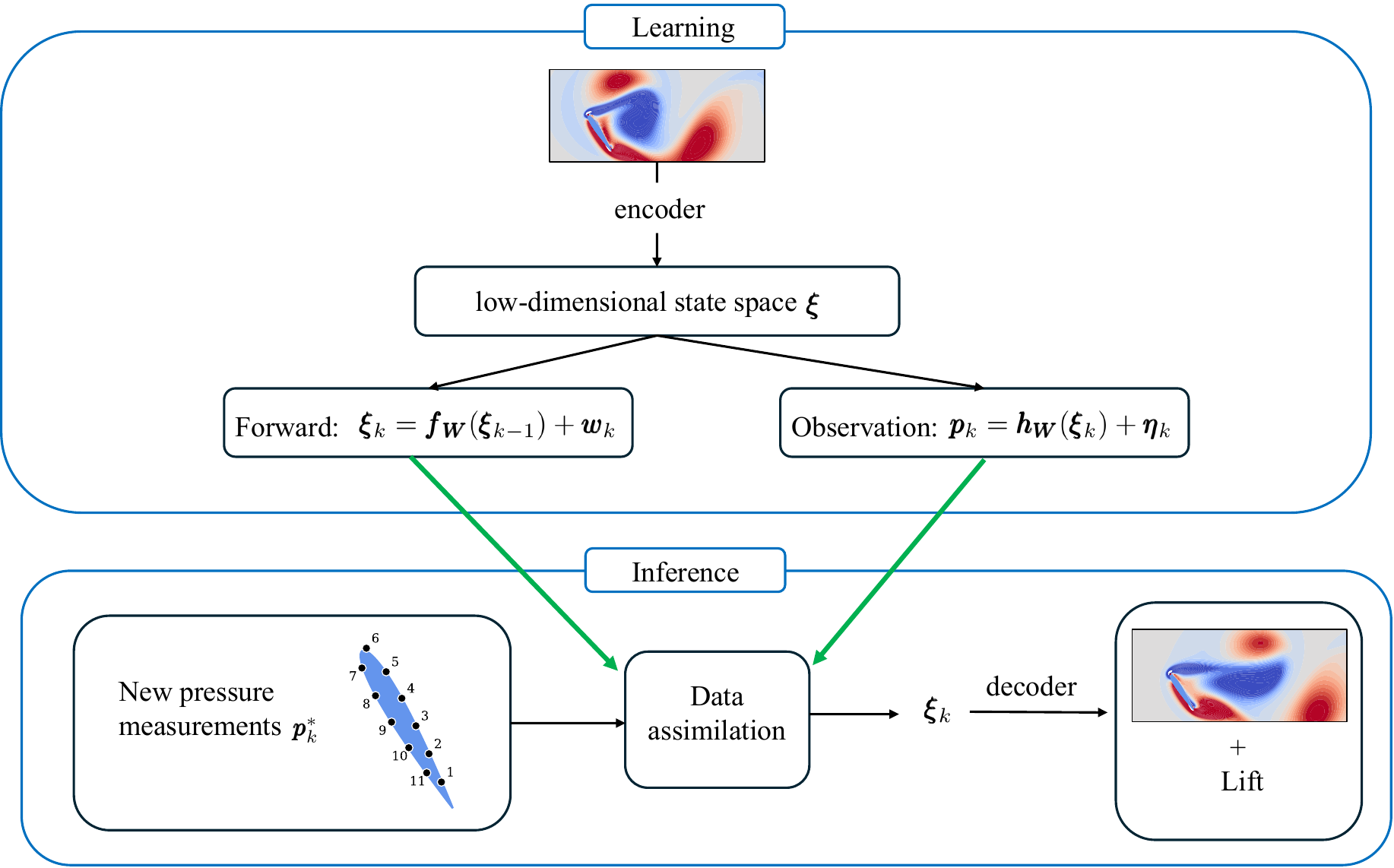}
\caption{\label{fig:diagram} Block diagram summarizing the framework developed and applied in this study.}
\end{figure*}

\begin{figure*}
\centering
\includegraphics[width=0.6\textwidth]{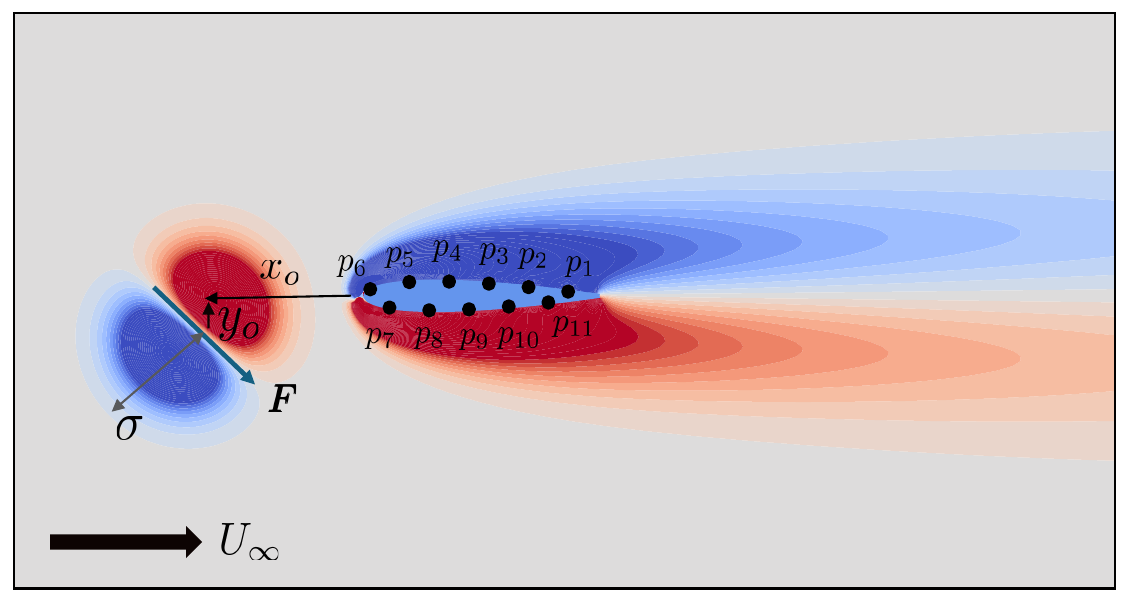}
\caption{\label{fig:configuration} Configuration of the problem, illustrating the relative position of the gust centre with respect to the leading edge of the airfoil, the size of the Gaussian disturbance and the indices of sensors mounted on the airfoil.}
\end{figure*}
In this study, we present a unified framework that integrates data-driven modeling with sequential estimation to reconstruct unsteady flow fields and lift from sparse, noisy surface pressure measurements. The central aim is to enable fast, robust, and uncertainty-aware inference under strongly disturbed conditions, where traditional simulation or filtering approaches become computationally prohibitive or inaccurate. Figure~\ref{fig:diagram} illustrates the overall concept. At the heart of the framework is a compact latent space, $\lat$, learned through nonlinear compression techniques based on DL. This reduced representation captures the essential flow physics while filtering out redundancy, thereby providing a tractable foundation for data assimilation. Surrogate dynamical and observation models---required by the DA task---are then trained directly in this latent space. During inference, new surface pressure data are assimilated with the EnKF. Finally, the updated latent samples are decoded back into the physical space, yielding flow-field reconstructions and load estimates along with their associated uncertainties. In this section, we describe the details of the generation of training data and the estimation framework itself.

\subsection{Problem statement}\label{problem-statement}
To generate training and testing data for DL tasks, we simulate an unsteady, incompressible two-dimensional flow over a NACA 0012 airfoil at fixed angles of attack (AoA), $\alpha \in \{20^\circ, 30^\circ, 40^\circ, 50^\circ, 60^\circ\}$. The undisturbed flow is characterized by a chord-based Reynolds number of $\mathrm{Re}=U_\infty c/\nu=100$, where $U_\infty$ is the freestream velocity, $c$ is the chord length, and $\nu$ is the kinematic viscosity of the fluid. To model disturbed conditions, we introduce localized gusts upstream of the airfoil at randomly selected instants during the natural vortex shedding cycle of the airfoil. These gusts are implemented as a body force $\force$---Gaussian distributed in space and time, and applied to the flow field---defined as:
\begin{equation}
    \force(x,y,t) = \rho U_\infty c^2 \frac{(D_x,D_y)}{\pi^{3/2} \sigma_x \sigma_y \sigma_t} \exp \left[ - \frac{(x-x_o)^2}{\sigma_x^2} - \frac{(y-y_o)^2}{\sigma_y^2} - \frac{(t-t_o)^2}{\sigma_t^2} \right],
\end{equation}
where $\rho$ is the fluid density, $(D_x,D_y)$ are the dimensionless forcing amplitudes in the streamwise and transverse directions, $\sigma_x$ and $\sigma_y$ are the spatial spreads of the forcing field, and $\sigma_t$ is its temporal width. The gust center is located at $(x_o,y_o)$ relative to the leading edge of the airfoil and introduced at time $t_o$. The schematic of the problem configuration is illustrated in figure~\ref{fig:configuration}. In all simulations, we fix $D_x=2.0$, $x_o=-0.8c$, while other gust parameters are randomly sampled from the ranges $D_y \in [-2.0,2.0]$, $\sigma_x=\sigma_y=\sigma \in [0.05c,0.2c]$, $y_o \in [-0.25c,0.25c]$, and $t_o \in [0.3c/U_\infty,5.0c/U_\infty]$. The upper bound for $t_o$ is intentionally selected near the vortex shedding period to ensure gusts interact with the airfoil at various phases within the first cycle. In this study, Latin Hypercube Sampling is employed to efficiently generate diverse combinations of gust parameters across the high-dimensional design space. This stratified sampling approach ensures uniform coverage of the parameter ranges while minimizing the total number of simulations required for training and evaluation. 

The direct numerical simulation of the incompressible Navier–Stokes equations is performed in vorticity–streamfunction form using the Lattice Green’s Function/Immersed Layers method developed by \citet{eldredge2022method}. This approach enables high-resolution flow computation within a relatively compact domain, made possible by the efficient treatment of boundary conditions via the lattice Green’s function. The simulations are conducted over a Cartesian domain spanning $(-2c,4c) \times (-2c,2c)$ in the $x$ and $y$ directions, respectively, with uniform grid spacing $\Delta x/c=0.02$. During the simulation, surface pressure $p_s$ relative to the ambient pressure $p_\infty$ is recorded at 11 predefined uniformly-distributed sensor locations (see figure~\ref{fig:configuration}). The pressure coefficient is computed using the standard definition:
\begin{equation}
    p = \frac{2(p_s - p_\infty)}{\rho U_\infty^2}.
\end{equation}
For training purposes, the vorticity data are extracted from a tighter region within the computational domain: $(-1.5c,3.3c) \times (-1.2c,1.2c)$ in the $x$ and $y$ directions, respectively. This subdomain is selected to maintain a balance between computational efficiency and physical fidelity, ensuring the full evolution of the gust disturbance and its interaction with the airfoil and wake are accurately captured. At each AoA, one undisturbed (baseline) case and 60 randomized gust cases are simulated, resulting in a dataset comprising 5 undisturbed cases and 300 disturbed (each with a single gust passage) cases in total. Each case is computed over 10 nondimensional convective time units, defined as $t=t^\prime U_\infty /c$, where $t^\prime$ is the dimensional time. This time horizon ensures that at least two full vortex shedding cycles are captured. Flow fields are sampled at a uniform time interval of $\Delta t=0.02$, yielding 500 snapshots per case. Altogether, the dataset consists of $152 \, 500$ flow snapshots (vorticity field, surface pressure at the designated sensor locations, and lift force) spanning a wide range of aerodynamic conditions. The flow states are defined as the vorticity field and the corresponding lift coefficient, which are to be estimated by assimilating synthetic surface pressure measurements.

\subsection{Finding the low-dimensional latent space}\label{sec:data-compression}
\begin{figure*}
\centering
\includegraphics[width=1.0\textwidth]{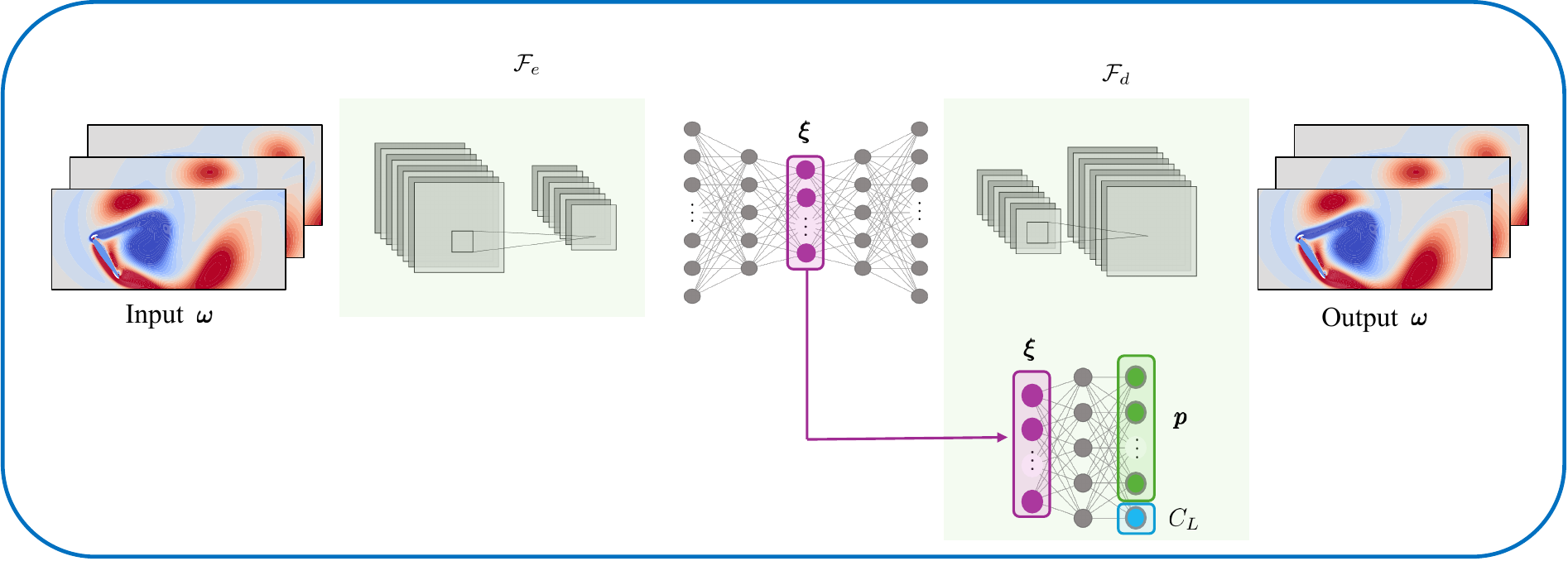}
\caption{\label{fig:network} Network architecture in the present study, including a non-linear physics-augmented convolutional autoencoder. The observation operator that maps the learned latent vector to the pressure measurements is learned on-the-fly while training the autoencoder.}
\end{figure*}
To enable efficient data assimilation and accurate flow reconstruction, we first construct a compact, low-dimensional representation of the high-dimensional flow state. This reduced-order space is designed to capture the dominant coherent structures in the vorticity field, along with the associated lift dynamics, and serves as the latent state space for sequential filtering. To achieve this, we extend the concept of the lift-augmented autoencoder introduced by \citet{fukami2023grasping} and adapt it to the requirements of our framework. The architecture of the resulting nonlinear physics-augmented convolutional autoencoder is illustrated in Figure~\ref{fig:network} and table~\ref{tab:network_blocks}. In this model, the high-dimensional vorticity field $\vor \in \mathbb{R}^l$ is encoded into a low-dimensional latent vector $\lat \in \mathbb{R}^n$ via a NN encoder $\encoder$. The latent representation $\lat$, with $n \ll l$, is then passed through a decoder $\decoder$ to reconstruct not only the original vorticity field but also the corresponding lift coefficient and surface pressure values at prescribed sensor locations. This design ensures that the latent space retains sufficient information relevant to both global flow structure and aerodynamic observables. 

\begin{table}[htbp]
  \centering
  \caption{Network architecture of the encoder, decoder, lift reconstruction, and pressure observation models. The activation function used is Tangent Hyperbolic.}
  \label{tab:network_blocks}
  \renewcommand{\arraystretch}{3.0}  
  \begin{tabularx}{\textwidth}{|Y|Y||Y|Y||Y|Y|}
    \hline
    \multicolumn{2}{|c||}{\textbf{Encoder}} & 
    \multicolumn{2}{c||}{\textbf{Decoder}} & 
    \multicolumn{2}{c|}{\textbf{Lift/Pressure Observation}} \\
    \hline
    \textbf{Layer} & \textbf{Data Size} & 
    \textbf{Layer} & \textbf{Data Size} & 
    \textbf{Layer} & \textbf{Data Size} \\
    \hline
    Input & (240, 120, 1) & Dense & (128) & Dense & (32) \\
     \hline
    \shortstack{Conv2D\\(3,3,32)} & (240, 120, 32) & Dense & (256) & Dense & (64) \\
     \hline
    \shortstack{Conv2D\\(3,3,32)} & (240, 120, 32) & Dense & (288) & Dense & (32) \\
     \hline
    \shortstack{MaxPooling2D\\(2,2)} & (120, 60, 32) & Reshape & (12, 6, 4) &\shortstack{Output 2\\(Pressure)} & (11) \\
     \hline
    \shortstack{Conv2D\\(3,3,16)} & (120, 60, 16) & \shortstack{Conv2D\\(3,3,4)} & (12, 6, 4) & \shortstack{Output 3\\(Lift)} & (1) \\
     \hline
    \shortstack{Conv2D\\(3,3,16)} & (120, 60, 16) & \shortstack{Conv2D\\(3,3,4)} & (12, 6, 4) & &  \\
     \hline
    \shortstack{MaxPooling2D\\(2,2)} & (60, 30, 16) & \shortstack{UpSampling2D\\(5,5)} & (60, 30, 4) &  &  \\
     \hline
    \shortstack{Conv2D\\(3,3,8)} & (60, 30, 8) & \shortstack{Conv2D\\(3,3,8)} & (60, 30, 8) &  &  \\
     \hline
    \shortstack{Conv2D\\(3,3,8)} & (60, 30, 8) & \shortstack{Conv2D\\(3,3,8)} & (60, 30, 8) &  &  \\
     \hline
    \shortstack{MaxPooling2D\\(5,5)} & (12, 6, 8) & \shortstack{UpSampling2D\\(2,2)} & (120, 60, 8) &  &  \\
     \hline
    \shortstack{Conv2D\\(3,3,4)} & (12, 6, 4) & \shortstack{Conv2D\\(3,3,16)} & (120, 60, 16) &  &  \\
     \hline
    \shortstack{Conv2D\\(3,3,4)} & (12, 6, 4) & \shortstack{Conv2D\\(3,3,16)} & (120, 60, 16) &  &  \\
     \hline
    Reshape & (288) & \shortstack{UpSampling2D\\(2,2)} & (240, 120, 16) &  &  \\
     \hline
    Dense & (256) & \shortstack{Conv2D\\(3,3,32)} & (240, 120, 32) &  &  \\
     \hline
    Dense & (128) & \shortstack{Conv2D\\(3,3,32)} & (240, 120, 32) &  &  \\
     \hline
    \shortstack{Dense\\(Latent vector)} & (7) & \shortstack{Output 1\\(Vorticity)} & (240, 120, 1) &  &  \\
     \hline
  \end{tabularx}
\end{table}

The network parameters $\weights$---comprising all weights and biases---are optimized by minimizing the following composite loss function with weight coefficients $\beta$:
\begin{align}\label{eq:loss}
    \weights = \underset{\weights}{\arg\min}(\mathcal{L}_{AE}) &= \underset{\weights}{\arg\min} \Big( 
    \underbrace{\beta_{\vor} ||\vor - \h{\vor}||_2^2}_{\text{vorticity reconstruction}} 
    + \underbrace{\beta_{\pres} ||\pres - \h{\pres}||_2^2}_{\text{pressure reconstruction}} \notag \\
    &+ \underbrace{\beta_{\lift} ||\lift - \h{C}_L||_2^2}_{\text{lift reconstruction}} 
    \ + \ \underbrace{\beta_{t} ||\lat_{t+1} - 2 \lat_{t} + \lat_{t-1}||_2^2}_{\text{temporal smoothness}} \Big).
\end{align}
Here, the lift and pressure reconstruction terms are incorporated into the loss function $\mathcal{L}_{AE}$ to regularize the learned latent space and ensure that it remains informative with respect to key aerodynamic quantities. Notably, this formulation learns the observation operator---mapping the low-dimensional latent states $\lat$ to surface pressure coefficient measurements, $\pres = \obs(\lat) \in \mathbb{R}^{d}$, with nonlinear activation function (tangent hyperbolic)---during the training process. Rather than training a separate network to learn the observation mapping post hoc, we integrate it directly into the autoencoder training. This joint learning approach has two key advantages: (1) it constrains the latent space to remain consistent with the pressure signatures, thus promoting sensor-relevant representations; and (2) it improves the accuracy and robustness of the learned observation operator while accelerating convergence during training. By embedding observability constraints (though partial) within the compression task, the resulting latent space not only captures the dominant flow structures but also preserves essential information required for downstream tasks such as state estimation and control. As a result, the decoder reconstructs the vorticity field and lift as the original flow states, as well as pressure observations.

Conventional autoencoders process each data snapshot independently, disregarding the temporal correlations inherent in a transient fluid flow. As a result, the learned latent manifold may not form a dynamically smooth or physically consistent phase space, leading to suboptimal compression and reduced suitability for learning time-evolving dynamics. However, for systems governed by smoothly evolving physical processes, such as transient fluid flow, the reduced latent space should ideally preserve this temporal quality. In other words, smooth evolution in the original state space should be mirrored by equally smooth trajectories in the latent representation to ensure physically meaningful and dynamically consistent encoding. To address this issue, we introduce a temporal regularization term that penalizes abrupt changes in the latent space from one input snapshot to the next in the sequence. Specifically, we enforce smoothness by minimizing the second time derivative of the latent variables. Compared to using a first derivative, this higher-order constraint has been shown to yield more stable and consistent representations during training \citep{xie2024smooth}. By discouraging sudden accelerations in the latent dynamics, this approach promotes smoother temporal evolution and mitigates spurious jumps, which is particularly important for sequential filtering and forecasting tasks. 

The loss weights $\beta_i$ in Eq.~\eqref{eq:loss}, where $i \in \{ \vor, \pres, \lift, t \}$, are used to balance the relative contributions of each loss component to the total objective. These weights are carefully tuned to emphasize accurate reconstruction of the vorticity field and surface pressure---both of which are critical for ensuring reliable state estimation via the observation operator---while keeping the lift and temporal smoothness terms at comparable but lower magnitudes. Specifically, the lift and temporal losses are scaled to be approximately one order of magnitude smaller than the dominant reconstruction losses to serve as regularizers without overpowering the main signal. Based on empirical testing, we set $\beta_{\vor}=1.0$, $\beta_{\pres}=100.0$, $\beta_{\lift}=1.0$, and $\beta_t=5000.0$. It is important to note that the relatively large coefficient $\beta_t$ does not indicate that the temporal regularization term dominates the overall objective. In practice, the discrete second-order temporal increments, $||\lat_{t+1} - 2 \lat_{t} + \lat_{t-1}||_2^2$, are several orders of magnitude smaller than the reconstruction losses. As a result, a larger weighting factor is required to balance their contribution during training. This choice of scaling allows the temporal smoothness constraint to meaningfully influence the optimization process while maintaining the primacy of the reconstruction objectives.

For training, $70 \%$ of the total random gust cases---each spanning the full gust–airfoil interaction duration---are used for model optimization, while the remaining $30 \%$ are held out for validation and testing. Separating training and test sets by distinct gust encounters---rather than by randomly selecting individual snapshots across all cases---better reflects realistic deployment scenarios. In practice, models are trained on a subset of gust conditions and are expected to generalize to entirely new gust patterns. Training is terminated when the validation loss converges and plateaus (around epoch $\approx 2000$), indicating stable learning.

\subsection{Sequential filtering}\label{sec:filtering}
Sequential filtering offers a principled framework for estimating the evolving state probability distribution of a dynamical system by continuously incorporating streaming observational data. This study enables online inference of latent aerodynamic states from sparse and noisy surface pressure measurements, facilitated by learned forecast and observation models discussed in Section \ref{sec:surrogate-models}. For a comprehensive overview of sequential filtering methods---including their foundations and applications in aerodynamic contexts---we refer the reader to the recent work by \citet{eldredge2025practical}.

Each sequential filtering algorithm consists of two key substeps: the \emph{forecast} (or prediction) step and the \emph{analysis} (or update) step. In the forecast step, the system states---here, the latent variables $\lat$---are advanced in time using a dynamical model $\pmb{f}$. When the governing dynamics are not known a priori, they can be learned from data using a NN, resulting in a learned forecast model $\frw$, parameterized by network weights $\weights$. The formulation assumes that the current state depends only on the immediate past state, thus preserving the Markov property of the underlying system dynamics within the reduced latent space. \rev{In the exact Bayesian framework, the forecast step provides an estimate of the prior distribution of states at that instant, $\pi(\lat_k|\pres_{1:k-1})$. Upon the arrival of new measurements, $\pres_k$, the analysis step updates the prior to the posterior distribution, $\pi(\lat_k|\pres_{1:k})$.} This Bayesian inference problem requires an observation operator $\pmb{h}$ that maps the latent state $\lat$ to observables (e.g., surface pressures), with additive observation noise $\obsnoise$. As with the forecast model, this observation operator can be learned from data using a NN, $\obs$, parameterized by its own set of weights.

In the classical setting, when both process and measurement noises are Gaussian and the forecast and observation operators are linear, the exact posterior distribution of the system states at each time step remains Gaussian and can be computed analytically. However, in aerodynamic applications such as the one considered in this study, the forecast and observation operators are highly nonlinear, including when they are learned through NNs with nonlinear activations. As a result, the posterior distribution becomes non-Gaussian, even if the noise distributions remain Gaussian. To address this, various extensions of the standard Kalman filter have been developed for nonlinear systems. Among them, the EnKF has proven particularly effective, especially in moderate- to high-dimensional spaces. Originally introduced by \citet{evensen1994sequential}, the EnKF \rev{approximates} the distribution of system states, $\pi(\lat_k|\pres_{1:k})$, using a finite ensemble of samples. It has been successfully applied to a wide range of problems, including in fluid dynamics and reduced-order modeling \citep{eldredge2025practical, da2018ensemble, moldovan2021multigrid, le2021ensemble}.

In the EnKF framework, an ensemble of latent states $\{\lat^i\}_{i=1}^M$ is drawn from some initial distribution. During the forecast step, each ensemble member is independently propagated forward in time using the learned dynamical model:
\begin{equation}\label{eq:forecast}
    \lat^i_{k|k-1} = \frw(\lat^i_{k-1}) + \frwnoise^i_k,
\end{equation}
to give a prior ensemble of states $\pmb{\Xi}_{k|k-1} \coloneq \{\lat^i_{k|k-1} \}_{i=1}^M$. In this equation $\frwnoise \sim \mathcal{N}(\pmb{0},\pmb{Q})$ represents the process noise and $\pmb{Q}$ is an empirical covariance, determined from the one-step residual differences between predicted and true latent trajectories averaged over time for all test cases. This approach is designed to balance the tradeoff between bias and variance in the ensemble of states.

The observation operator is then used to calculate the predicted observation associated with each ensemble member:
\begin{equation}\label{eq:observation}
    \pres^i_k = \obs(\lat^i_{k|k-1}) + \obsnoise^i_k,
\end{equation}
where $\obsnoise \sim \mathcal{N}(\pmb{0}, \obsCov)$ denotes the observation noise with covariance $\obsCov$. In this context, its standard deviation (square root of diagonal elements in $\obsCov$) is fixed for all sensors and set equal to the mean observation error computed from the learned observation operator on the test cases (except in some experiments in which the noise of selected sensors is increased to simulate sensor failure). This choice models the observation uncertainty as the combined effect of sensor noise and observation surrogate approximation error. Collectively, the ensemble of noisy predicted observations is denoted by $\pmb{P}_k \coloneq \{\pres^i_k \}_{i=1}^M$.

When a new observation becomes available, each ensemble member is updated during the analysis step using the Kalman update formula:
\begin{equation}\label{eq:kalman_update}
    \lat^i_k = \lat^i_{k|k-1} + \gain_k \left( \pres^*_k - \pres_k^i \right),
\end{equation}
where $\pres_k^*$ denotes the synthetically generated noisy measurements obtained from CFD data by adding an i.i.d. random noise term $\obsnoise_k$. The term in parentheses is referred to as the \emph{innovation}---the discrepancy between the measured and predicted observations. \rev{The Kalman gain $\gain_k$ determines how much weight is given to the new observation relative to the predicted state and is computed as
$\gain = \stateObsCov \, \obsCrossCov^{-1}$, where $\stateObsCov$ denotes the cross-covariance between the state and observation ensembles, and $\obsCrossCov$ represents the covariance of the noise-injected predicted observations \mbox{\citep{asch2016data}}. These covariances are approximated empirically from the ensemble in accordance with \mbox{\citet{asch2016data}}, with details provided in \mbox{Appendix~\ref{sec:apkalman}}.} This update strategy enables the EnKF to provide an efficient and robust posterior estimate, even in complex, nonlinear, and partially observed systems.

\rev{At each assimilation step, the dominant directions along which measurements correct the latent state are identified using Gramian operators constructed from the covariance-weighted Jacobian of the observation operator, $\nabla \obs$. The eigenvectors of the resulting state-space and observation-space Gramians characterize the dominant correction directions during the analysis step of the EnKF \mbox{\citep{le2021low}}.}
Following \citet{cui2021data}, in a nonlinear Gaussian setting, the correction directions in the state space can be identified via the state-space Gramian, defined as:
\begin{equation}\label{eq:state-Gramian}
    \stateGram = \int \left( \obsCov^{-1/2} \nabla \obs(\lat) \stateCov^{1/2} \right)^{\top} \left( \obsCov^{-1/2} \nabla \obs(\lat) \stateCov^{1/2} \right) \mathrm{d} \pi(\lat) \in \mathbb{R}^{n \times n},
\end{equation}
where the expectation is taken with respect to the prior distribution. Here, $\nabla \obs(\lat)$ denotes the Jacobian of the observation operator, computed via automatic differentiation within the NN framework, and $\stateCov$ is the prior state covariance matrix. By construction, $\stateGram$ is positive semi-definite, and its eigendecomposition $\stateGram=\stateMode \stateEig^2 \stateMode^{\top}$ yields an orthonormal basis $\stateMode \in \mathbb{R}^{n \times n}$ for the state space, with corresponding eigenvalues $\stateEig \in \mathbb{R}^{n \times n}$. In a similar manner, the observation-space Gramian $\obsGram$ was defined by \citet{le2022low} as
\begin{equation}\label{eq:observation-Gramian}
    \obsGram = \int \left( \obsCov^{-1/2} \nabla \obs(\lat) \stateCov^{1/2} \right) \left( \obsCov^{-1/2} \nabla \obs(\lat) \stateCov^{1/2} \right)^{\top} \mathrm{d} \pi(\lat) \in \mathbb{R}^{d \times d}.
\end{equation}
Like its state space counterpart, $\obsGram$ is positive semi-definite, and its eigendecomposition $\obsGram=\obsMode \obsEig^2 \obsMode^{\top}$ yields an orthonormal basis $\obsMode \in \mathbb{R}^{d \times d}$ for the observation space, with corresponding eigenvalues $\obsEig \in \mathbb{R}^{d \times d}$. Both Gramians can be efficiently approximated via the Monte Carlo method over the prior ensemble. \rev{In practice, these eigenvalues are sorted in descending order, so the leading eigenvectors in $\obsMode$ correspond to the most informative measurement directions. These directions contribute most strongly to correcting the corresponding leading latent eigendirections in $\stateMode$. These dominant directions enable us to analyze the informativeness of the sensors during the filtering update step.
It should be noted that the trailing (small eigenvalue) state and observation modes, although contributing minimally to the measurement update step, may still play an important role in reconstructing the high-dimensional vorticity field, since all latent variables participate in the decoding process.}

\subsection{Learning surrogate models in the latent space}\label{sec:surrogate-models}
As described in Section~\ref{sec:filtering}, the two essential components for the sequential filtering are the forecast and observation operators. We have already learned a surrogate observation operator, $\pres = \obs(\lat)$, by virtue of the side network in our physics-augmented convolutional autoencoder architecture. With the low-dimensional latent states $\lat$ established via the autoencoder, we now aim to learn the temporal evolution of the flow state in this reduced latent space. These surrogate models replace traditional high-fidelity solvers by approximating the forecast and observation operators directly in the reduced space, enabling fast and efficient DA. For a comprehensive overview of various approaches to learning surrogate models for dynamical systems, readers are referred to \citet{bach2024inverse}.

\begin{figure*}
\centering
\includegraphics[width=0.6\textwidth]{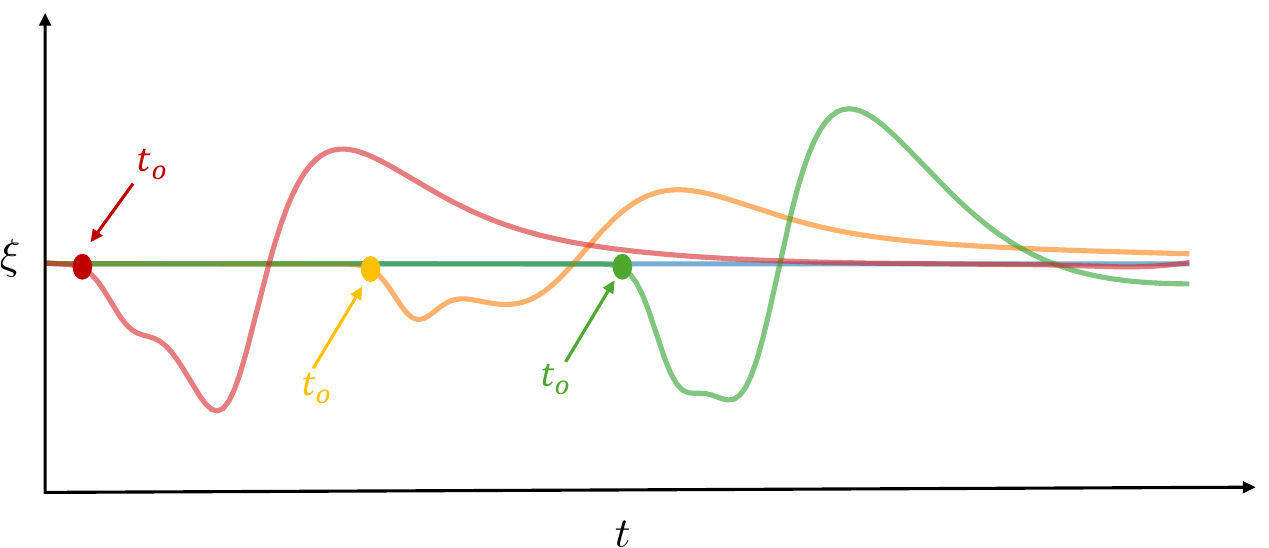}
\caption{\label{fig:bifurcation} Bifurcation in learned latent trajectories for various gust-encounter aerodynamic cases, with each trajectory shown in a different color. For each case, the latent trajectory deviates from its corresponding undisturbed path (base flow with the airfoil at $\alpha=20^\circ$) at the gust onset time $t_o$, reflecting the system's response to external disturbance. All trajectories are obtained directly from the encoded data using the trained autoencoder, without any data assimilation or neural ODE.}
\end{figure*}

Sequential DA requires a transition model that governs the temporal evolution of the system states. As discussed in Section~\ref{sec:introduction}, filtering methods such as the Kalman filter are derived under the assumption of Markovian dynamics, where the future state depends solely on the current state. To model latent dynamics in a Markovian fashion, we employ a Neural ODE framework that learns a continuous-time dynamical system in the latent space, expressed as:
\begin{equation}\label{eq:NODE}
    \frac{\mathrm{d} \lat}{\mathrm{d}t} = \frww(\lat),
\end{equation}
where $\frww$ is a NN parametrized by $\weights$. This approach offers several advantages over traditional discrete-time models: it allows for variable time intervals, supports evaluation at arbitrary time points, and enables adaptive time integration strategies. To recover a standard discrete-time state transition model required by filtering, we discretize Eq.~\eqref{eq:NODE} using a first-order forward Euler scheme:
\begin{equation}
    \lat_{k} = \frw \left(\lat_{k-1} \right),
\end{equation}
where $\frw \left(\lat_{k-1} \right) = \frww \left(\lat_{k-1} \right) \Delta t + \lat_{k-1}$ with $k$ the time index. \rev{Although a Neural ODE is adopted here to model the latent dynamics, the proposed framework is not restricted to this choice. Any alternative dynamical model capable of propagating the latent state forward in time can be incorporated within the same sequential filtering structure, provided it can describe this dynamical update at a wide variety of states. An RNN-based approach is also likely suitable, for example.}

An illustration of latent trajectories under different random gust conditions is shown in Figure~\ref{fig:bifurcation}. Prior to the introduction of the gusts, all trajectories closely follow the undisturbed path corresponding to the baseline case. In Figure~\ref{fig:bifurcation}, the baseline is flat, corresponding to a steady flow with a time-invariant point in the latent space. However, most cases considered in this work will have time-periodic baseline trajectories reflecting the bluff-body vortex shedding of a wing at a significant angle of attack. When the gust is introduced at a randomly selected time (denoted by solid circles in the figure) within the shedding cycle, the corresponding trajectory deviates from the baseline and subsequently returns to the baseline path once the gust has convected sufficiently far from the airfoil. This bifurcation behavior captures the transient influence of the gust on the flow dynamics. \rev{The Neural ODE does not contain information about gust timing; therefore, trajectory separation is triggered by assimilation. Once the assimilated surface measurements detect the onset of the disturbance and steer the latent state toward the correct branch, the forecast operator is then expected to propagate the state along the subsequent disturbed trajectory (with help from ongoing measurement updates to maintain this trajectory).} Motivated by this behavior, and to ensure that the Neural ODE accurately captures the disturbed dynamics under diverse disturbance conditions across aerodynamic cases, the training snapshots for each disturbed case are restricted to begin 10 time steps (10$\Delta t$) after the gust onset, and encompassing the full gust–airfoil interaction that ensues. This targeted training strategy enables the Neural ODE to learn the post-bifurcation dynamics effectively with minimal one-step error, ensuring that during sequential filtering it can reliably propagate the corrected latent state together with the measurement update. 

\rev{Conditioned on the encoded angle of attack (selected from five discrete AoA values) and the initial latent state, the forecast model is constructed to evolve the flow dynamics during the approach and interaction of a disturbance with the fixed-angle airfoil.} The network architecture is summarized in Table~\ref{tab:network_blocks_forecast} and is trained using both undisturbed and gust-disturbed cases.

\begin{table}[htbp]
  \centering
  \caption{Network architecture of the Neural ODE. The activation function used is Tangent Hyperbolic.}
  \label{tab:network_blocks_forecast}
  \renewcommand{\arraystretch}{2.0}  
  \begin{tabularx}{0.5\textwidth}{|Y|Y|}
    \hline
    \multicolumn{2}{|c|}{\textbf{Neural ODE}} \\
    \hline
    \textbf{Layer} & \textbf{Data Size} \\
    \hline
    \shortstack{Input\\(Latent Vector)\\+\\(Encoded AoA)} & \shortstack{(7)\\+\\(5)} \\
     \hline
    Dense & (128) \\
     \hline
    Dense & (256) \\
     \hline
    Dense & (128) \\
     \hline
    \shortstack{Dense\\(Latent Vector)} & (7) \\
     \hline
  \end{tabularx}
\end{table}

In most applications, Neural ODEs are trained by minimizing a one-step prediction error. However, this approach has a well-known limitation: the forward model is trained using teacher forcing on one-step losses, so during inference, when the model is rolled out autoregressively, prediction errors accumulate over time, leading to degradation in accuracy. To mitigate this issue, we also include a loss term that penalizes prediction error over a long-time prediction horizon. Specifically, the model parameters $\weights_f$ are obtained by solving:
\begin{equation}
    \weights_{f} =  \underset{\weights_{f}}{\arg\min} \Big( \underbrace{ \beta_{f,1} ||\lat - \h{\lat}||_2^2}_{\text{rollout loss}} + \underbrace{\beta_{f,2}  ||\lat_{1:T} - \frw(\lat_{0:T-1})||_2^2}_{\text{one-step loss}} \Big),
\end{equation}
where $\beta_{f,1}$ and $\beta_{f,2}$ are weighting factors assigned to the rollout and one-step losses, respectively.
To emphasize accurate prediction during the initial phase of gust–airfoil interaction, we decay the rollout weighting over time as $\beta_{f,1} = 1 - 0.5(t-1)/(T-1) $ for $t=1, \cdots, T$, with $T$ denoting the final rollout time step. This schedule ensures that higher weight is given to early predictions during disturbance events and gradually reduces the contribution of later predictions within the rollout horizon. In this formulation, we set $\beta_{f,1} = 1.0$ and $\beta_{f,2} = 150.0$, as it was observed that these weights give better predictions.

With the surrogate models, as well as the process noise and observation noise covariances, now learned in the reduced-order space, all necessary components for sequential DA are in place. Note that after evolving ensembles forward through the deterministic learned forecast operator $\frw$, an i.i.d. random process noise term $\frwnoise_k$ is added to each ensemble member to represent model uncertainty. \rev{The measurement update acts as a mechanism that identifies departures from the undisturbed baseline trajectory and redirects the state estimate toward the correct disturbed branch. Thus, to ensure timely detection of disturbance onset, data assimilation is performed at every time step, yielding 500 assimilation steps per case over approximately two vortex-shedding cycles (about 250 steps per shedding period). The time step $\Delta t = 0.02$ is used consistently for both state evolution and measurement assimilation. Notably, this assimilation frequency remains lower than the operating frequency of practical pressure sensors, underscoring the feasibility of the proposed framework.}

\section{Results and discussion}\label{sec:results}
In this section, we first focus on the compression of full-dimensional flow data into a reduced-dimensional latent space via a physics-augmented convolutional autoencoder. Then, we discuss the training of surrogate models necessary for filtering. Finally, we present results for data assimilation in the latent space.

\subsection{Construction of latent space and learned operators}
 
Previous researchers have demonstrated that high-dimensional disturbed aerodynamic flows can be effectively projected onto very low-dimensional manifolds---as low as three-dimensional---using nonlinear autoencoders \citep{fukami2023grasping, mousavi2025low, fukami2025extreme}. However, it is important to note that when additional regularization terms---such as temporal smoothness and pressure reconstruction losses---are incorporated into the training objective, the dimensionality of the latent space must generally be increased to compensate for the stricter constraints imposed on the encoded representation. This increase provides the necessary degrees of freedom to adequately capture the dominant flow features and enhance their observability within the compressed subspace.

\begin{figure*}
\centering
\includegraphics[width=0.5\textwidth]{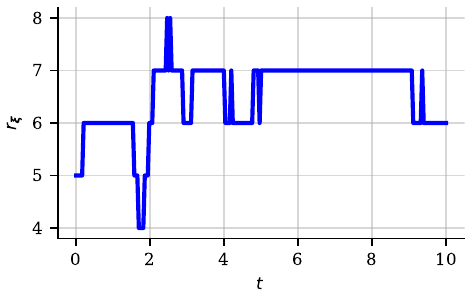}
\caption{\label{fig:latent_dim} The history of state rank $\stateRank$ for $n=10$. \rev{The first $\stateRank$ eigenvalues of $\stateGram$ retain $99 \%$ of cumulative energy.} The flow condition is ($\alpha=60^\circ$, $D_y=-0.71$, $\sigma=0.12$, $y_o=-0.19$, $t_o=3.3$).}
\end{figure*}

To determine an appropriate latent-space dimension $n$, we take the following systematic approach, guided both by the error trends in training the autoencoder and the effective state rank of the sequential filtering in the \rev{eigenspace}. We first conducted a systematic analysis using the network architecture shown in figure~\ref{fig:network} and table \ref{tab:network_blocks}. Specifically, we examined the validation loss as a function of latent dimension and observed that the loss plateaued beyond $n=7$, indicating diminishing returns in reconstruction fidelity with further increases in dimensionality. This finding is corroborated by a complementary experiment in which the dataset was compressed into a 10-dimensional latent space, followed by sequential DA and tracking the first $\stateRank$ eigendirections with $99 \%$ cumulative \rev{spectral energy}. As illustrated in figure~\ref{fig:latent_dim}, rarely did more than seven directions in the latent space exhibit sensitivity to variations in the sensor measurements, further supporting the choice of $n=7$. While it is possible to adopt a higher-dimensional latent representation, the goal in this work is to encode the high-dimensional flow field into the most compact latent space that still preserves the information necessary for accurate state estimation. Conversely, selecting a latent dimension smaller than seven risks excessive compression, compromising the uniqueness and identifiability of the latent representation of flow trajectories required for robust sequential filtering and inference. Overall, the choice of latent dimension reflects a tradeoff among reconstruction accuracy, compactness of the representation, and degree of identifiability required for reliable sequential estimation. Based on these analyses, we set $n=7$ as the latent dimension in this study.

\begin{figure*}
\centering
\includegraphics[width=1.0\textwidth]{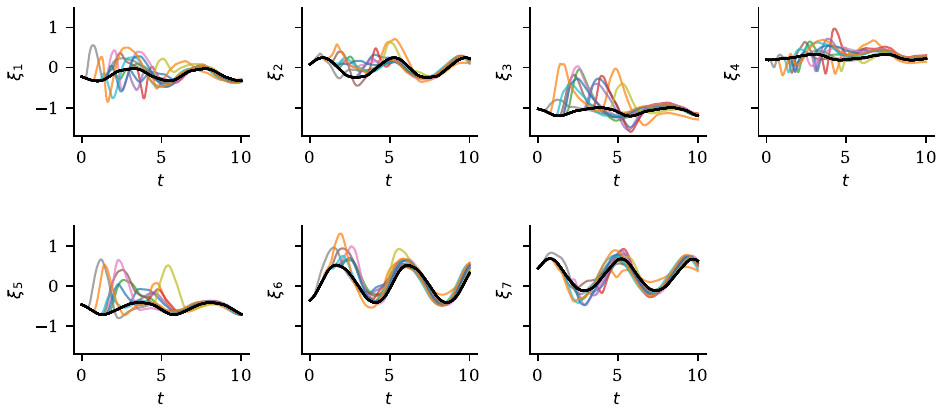}
\caption{\label{fig:latent} Latent trajectories for several disturbed aerodynamic cases. Each color corresponds to a different random gust introduced at a random time within the vortex shedding cycle, interacting with the airfoil at an AoA of $\alpha=40^\circ$, while the black curves correspond to the undisturbed case.}
\end{figure*}

To explore the extent to which the latent dynamics are interpretable, figure~\ref{fig:latent} presents the histories of the latent vector components for several independent disturbed cases at $\alpha=40^\circ$. The undisturbed flow over the airfoil at a high AoA (such as the case in this figure represented by black curves) exhibits characteristic periodic behavior, which is also clearly reflected in the latent space representation. The time window shown spans slightly more than two full vortex shedding cycles. The bifurcation highlighted earlier in figure~\ref{fig:bifurcation} is evident here as well, manifesting as divergence in the latent component histories due to the gust encounters at random phases of the shedding cycle. While these trajectories implicitly encode the influence of gust characteristics, vortex dynamics, and their interactions, the mapping between individual latent variables and specific physical phenomena remains difficult to disentangle. Determining which components of the latent vector correspond to particular aspects of the flow or external disturbances would require targeted analyses beyond the scope of this work. A systematic study aimed at interpreting these latent representations and rigorously connecting them to the underlying flow physics (e.g. \citep{smith2024cyclic}) remains an important direction for future research.

\begin{figure*}
\centering
\includegraphics[width=1.0\textwidth, trim=0 0 0 4.8, clip]{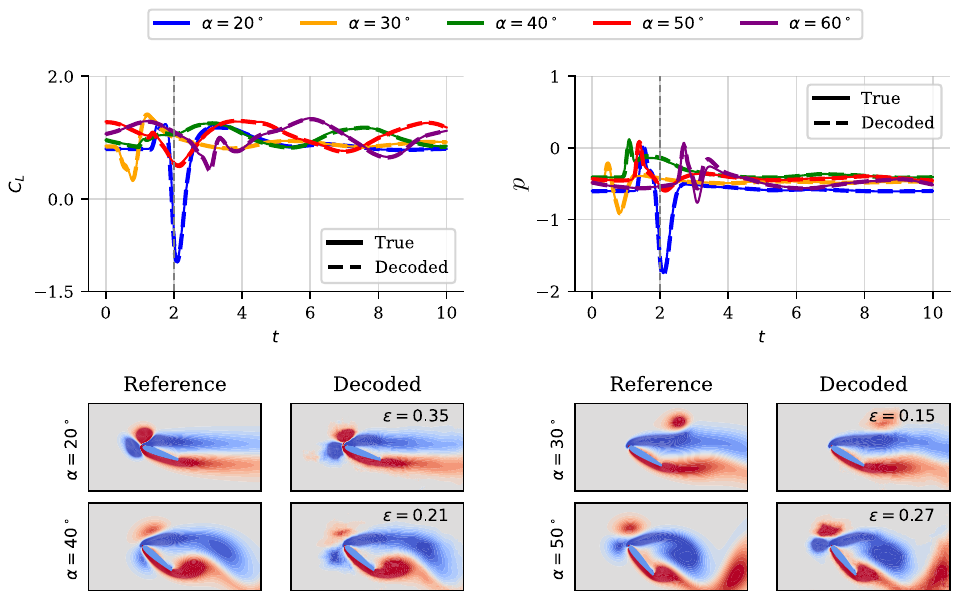}
\caption{\label{fig:decoded_lift_vor_pres} Comparison between the decoded and reference lift and pressure coefficients and vorticity fields for test cases under the following conditions: ($\alpha=20^\circ$, $D_y=-1.9$, $\sigma=0.1$, $y_o=0.22$, $t_o=1.5$); ($\alpha=30^\circ$, $D_y=0.44$, $\sigma=0.08$, $y_o=0.19$, $t_o=0.5$); ($\alpha=40^\circ$, $D_y=0.24$, $\sigma=0.19$, $y_o=-0.07$, $t_o=1.1$); ($\alpha=50^\circ$, $D_y=-0.64$, $\sigma=0.2$, $y_o=0.16$, $t_o=1.4$). The pressure plot corresponds to sensor 7. All vorticity plots are shown at time $t=2$, as indicated by the vertical dashed lines in the upper panel plots.}
\end{figure*}

The reconstruction of lift, pressure, and vorticity fields is achieved through the decoder part of the physics-augmented convolutional autoencoder. The performance of the decoder can be evaluated by comparing the decoded fields against the reference input fields. Figure~\ref{fig:decoded_lift_vor_pres} illustrates the results for five test cases under different AoA, showing reconstructed vorticity, lift coefficient, and one sensor's surface pressure coefficient for each case.
As evident in this figure, the reconstruction of the lift and the pressure coefficients (i.e., the learned observation operator) exhibit excellent performance across all four test cases. It is also clear that the presence of a disturbance causes the lift and pressure signals to deviate from their undisturbed periodic trajectories, returning to these trajectories once the gust convects downstream from the airfoil. The stronger the disturbance, the greater is the deviation of the lift and surface pressure readings from the baseline periodic path. The performance of the vorticity decoder, quantified by the reconstruction error, $\varepsilon = ||\vor - \h{\vor}||_2/||\vor||_2$, is overall satisfactory, especially considering the compact latent representation used in this study. As expected, the reconstruction error increases when a disturbance is present in the domain. The average pressure observation error over the whole test cases is used to determine the standard deviation of the random observation noise $\obsnoise$ in Eq.~\eqref{eq:observation}, in order to balance the bias and variance. 

\begin{figure*}
\centering
\includegraphics[width=0.5\textwidth]{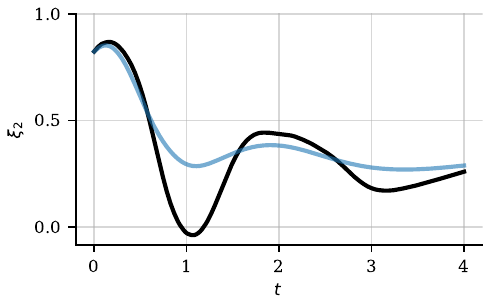}
\caption{\label{fig:forecast_error} Example of the history of true (black) and predicted (blue, from the learned forecast operator) latent state for the airfoil at $\alpha = 20^\circ$ during a gust encounter. \rev{Nevertheless, the one-step prediction required by DA remains highly accurate.}}
\end{figure*}

Now that we have demonstrated successful state compression in tandem with an accurately learned observation operator, there remains one more verification step before proceeding to DA: assessing the performance of the learned forecast operator described in Section~\ref{sec:surrogate-models}. Figure~\ref{fig:forecast_error} illustrates the history of one predicted latent variable, $\xi_2$, during gust passage in one of the test cases. Starting from the true initial state of a disturbed path, the autoregressive prediction initially tracks the true trajectory but eventually diverges. At first glance, this deviation may seem unexpected, given that the model was explicitly trained with a rollout loss designed to prevent error accumulation over long prediction horizons.
It is important to emphasize, however, that the one-step prediction remains highly accurate, as is essential for DA applications. Indeed, the learned forecast operator provides reliable short-term estimates during rollout predictions, ensuring that sequential updates in filtering can be based on precise time-local dynamics, even if longer-term rollout trajectories have larger errors.
Despite the larger prediction error observed in this example, the inclusion of the rollout loss does stabilize the network: the error does not grow without bound over time. Rather, the discrepancy arises because, while the original high-dimensional dynamics are Markovian, this quality is not necessarily preserved when compressed into a low-dimensional latent space. In other words, two nearby initial points in the compressed representation can evolve into significantly different trajectories.
This important limitation has been rigorously established in the foundational work of \citet{zwanzig1961memory} and \citet{mori1965transport}. According to Mori–Zwanzig theory, the evolution of any reduced subspace can be decomposed into a Markovian term, a memory (history-dependent) term, and an unobservable (noise) term. As demonstrated by \citet{ruiz2024benefits}, explicitly modeling the memory term can significantly improve predictive performance.
Following this insight, one promising way to improve the forecast operator is to incorporate history dependence. However, the philosophy of our current approach is that the responsibility for correcting deviations lies with the measurement update of each filter step. Thus, we deliberately choose to retain this imperfect, but short-term-accurate, forecast operator to evaluate the performance of our filtering framework under the realistic scenario of a non-ideal long-term predictive model. \rev{We emphasize that one could also adopt a different strategy---using, e.g., an RNN---and likely achieve similar estimation performance, given the forecast operator's limited role in this framework. Any other strategy would be similarly challenged by the lack of information about the gust arrival, and thus require the use of frequent measurement updates to detect this arrival.}

\subsection{Sequential filtering examples}

Now, everything is in place to set up the DA task. Pressure measurements from 11 sensors uniformly mounted on the airfoil (see figure~\ref{fig:configuration}) in CFD simulations with added noise are used for state estimation via the EnKF framework. Each member of the ensemble of estimated latent states is then propagated through the pre-trained decoder of the network architecture shown in figure~\ref{fig:network} to reconstruct both the lift coefficient and the vorticity field. For all cases in this section, a convergence study was performed, and we selected an ensemble size of $M=200$. \rev{We chose to use a large ensemble size to ensure that ensemble statistics were fully and unambiguously converged. The large ensemble is tractable because of the modest dimensions of the latent and measurement spaces.} \rev{The ensemble members are initialized from a Gaussian distribution whose mean is intentionally offset from the true latent state corresponding to the undisturbed initial flow over the airfoil. Specifically, the initial ensemble mean is biased by 0.5 relative to the truth, with a variance of 0.25, ensuring that the prior distribution is sufficiently separated from the true state and that convergence is nontrivial.} \rev{The computational efficiency of the proposed framework is demonstrated by performing 500 sequential forecast–analysis cycles over ten convective time units in approximately two seconds on a single GPU. This corresponds to an average runtime of roughly 4 milliseconds per assimilation cycle, demonstrating that the proposed framework is computationally lightweight and suitable for real-time implementation for practical sensor sampling rates.}

\begin{figure}[tb]
\centering
\includegraphics[width=1.0\textwidth]{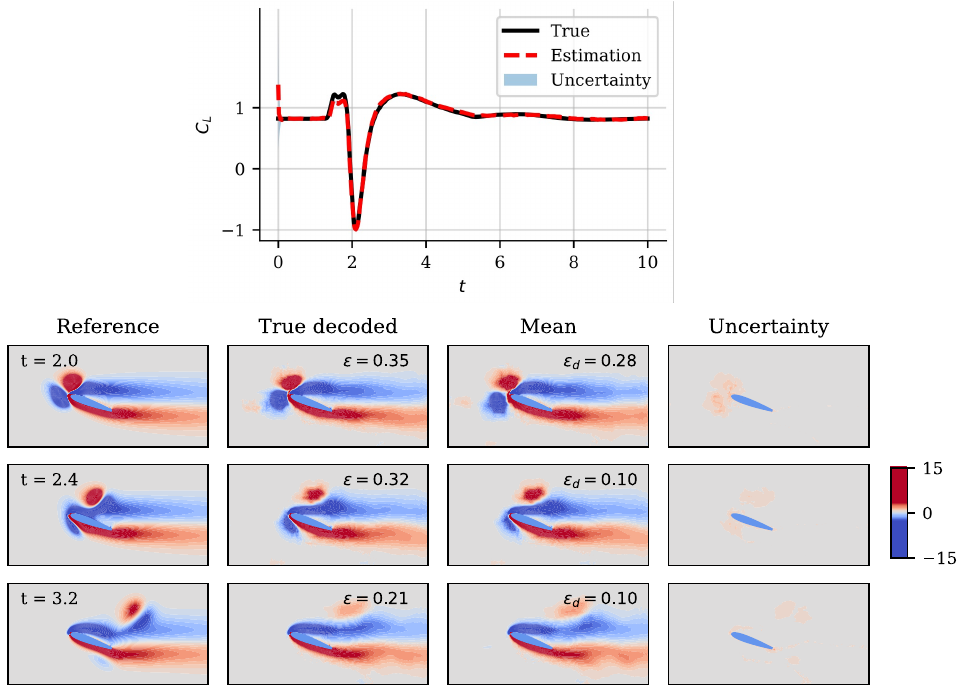}
\caption{\label{fig:estimation_AoA20} Estimated lift and vorticity field for disturbed aerodynamics at the condition ($\alpha=20^\circ$, $D_y=-1.9$, $\sigma=0.1$, $y_o=0.22$, $t_o=1.5$). The predicted ensemble mean, along with the $95\%$ confidence interval, is shown for the lift coefficient and the vorticity field.}
\end{figure}

\begin{figure}[tb]
\centering
\includegraphics[width=1.0\textwidth, trim=0 0 0 5, clip]{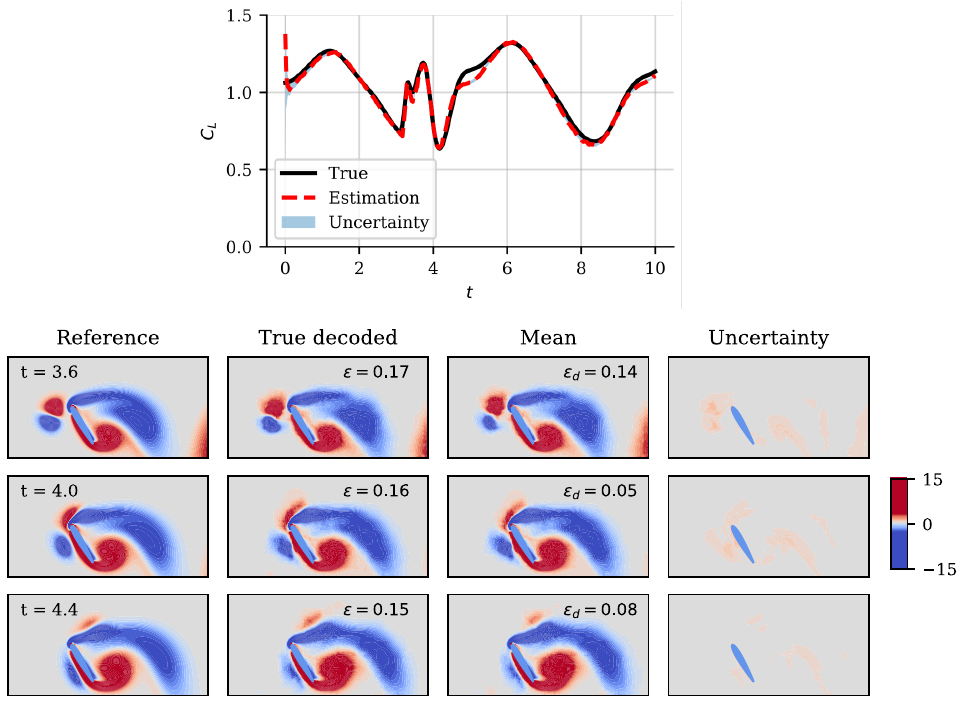}
\caption{\label{fig:estimation_AoA60} Estimated lift and vorticity field for disturbed aerodynamics at the condition ($\alpha=60^\circ$, $D_y=-0.71$, $\sigma=0.12$, $y_o=-0.19$, $t_o=3.3$). The predicted ensemble mean, along with the $95\%$ confidence interval, is shown for the lift coefficient and the vorticity field.}
\end{figure}

The results are presented in figures~\ref{fig:estimation_AoA20} and \ref{fig:estimation_AoA60} for the smallest and largest AoA considered in this study, namely $\alpha=20^\circ$ and $60^\circ$, respectively. In these test cases, disturbances with random parameters (strength, size, and orientation) are introduced at arbitrary instants within the first cycle. For both cases, the predicted lift coefficient shows excellent agreement with the true values throughout the assimilation horizon. Even when the initial ensemble distribution deviates substantially from the true initial state, the filter rapidly converges toward the correct trajectory and subsequently tracks the temporal variations induced by gust disturbances with high fidelity. \rev{This confirms the estimator's insensitivity to the initial condition.} The associated uncertainty bounds remain tight throughout the evolution, except during the initial DA phase when the ensemble is introduced to be far from the true state with relatively large variance. \rev{This demonstrates that the assimilation effectively leverages sparse pressure measurements to correct the latent state estimate---most critically by identifying and capturing the transition to the disturbed trajectory---even when the forecast operator alone exhibits substantial long-term prediction error and is incapable of anticipating the shift from the undisturbed baseline path to the gust-driven response.}

In figures~\ref{fig:estimation_AoA20} and \ref{fig:estimation_AoA60}, the estimated ensemble statistics are compared against the decoded vorticity reconstructed from the true latent vectors, referred to as the \emph{true decoded} vorticity. It is important to note that even the true decoded vorticity exhibits non-negligible reconstruction error relative to the reference field, and this error defines the attainable lower bound on prediction error. This field represents the best reconstruction achievable by the learned autoencoder. Accordingly, the prediction mean is evaluated relative to the true decoded field using the error metric $\varepsilon_d = ||\vor_d - \h{\vor}||_2/||\vor_d||_2$, where $\vor_d$ denotes the true decoded vorticity.
The vorticity reconstructions demonstrate similar trends: in all snapshots, the estimated ensemble mean fields are in good agreement with the true decoded vorticity, capturing the primary vortical structures and their evolution over time. As expected, reconstruction errors $\varepsilon_d$ are largest immediately after the disturbance enters the domain and hits the tip of the airfoil, but progressively decrease as the estimator assimilates new observations and corrects the predicted state. However, the gust core, as well as regions farther from the pressure sensors and near the periphery of the wake, show larger deviations from the reference flow. This discrepancy can be attributed to the lower sensitivity of the pressure sensors to flow variations in those areas, as reflected in the increased uncertainty of the estimates. Overall, the estimator remains highly confident (with narrow uncertainty bounds) and accurate in reconstructing the dominant flow features associated with both the disturbance and other large-scale vortical structures. 

\begin{figure}[tb]
\centering
\includegraphics[width=1.0\textwidth, trim=0 0 0 0, clip]{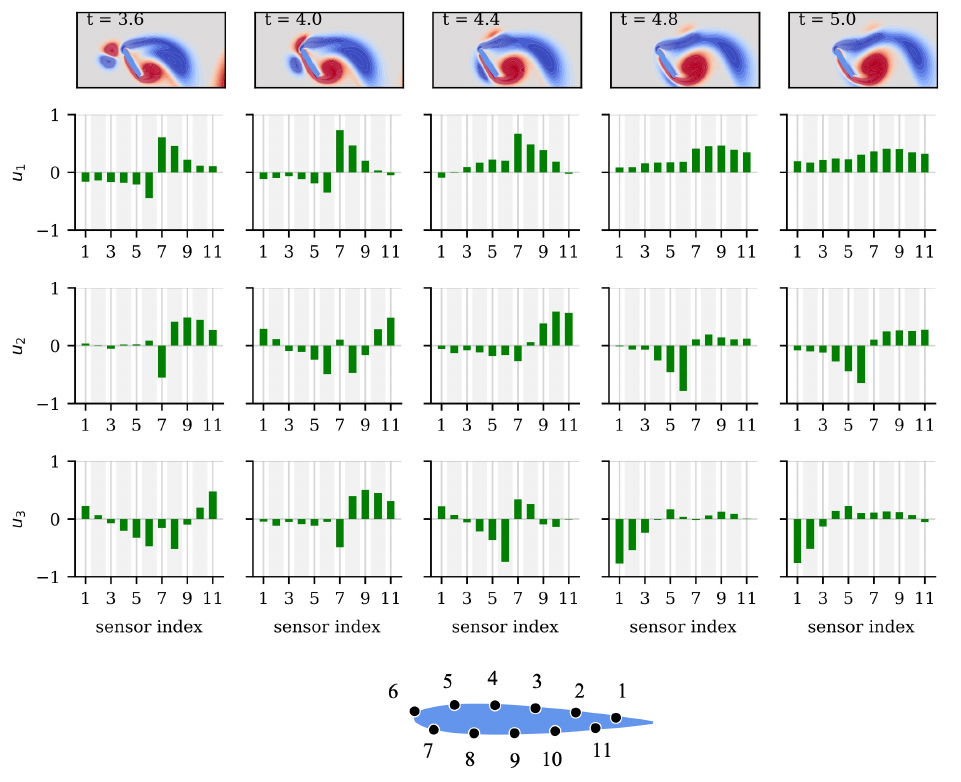}
\caption{\label{fig:obs_modes_AoA60} The first three observation modes at five distinct time instances during the gust-airfoil interaction for the case ($\alpha=60^\circ$, $D_y=-0.71$, $\sigma=0.12$, $y_o=-0.19$, $t_o=3.3$). How the sensor numbers are ordered in the eigenvector plots is shown on the airfoil.}
\end{figure}

The \rev{eigendeocmposition of the observation-space Gramian enables} a systematic analysis of how and which measurement sensors contribute most significantly to the update step. It is thus insightful to examine the observation modes corresponding to the sensor configuration and their role in informing state estimation. Figure~\ref{fig:obs_modes_AoA60} presents the leading observation eigenmodes---specifically $u_1$, $u_2$, and $u_3$. The sensor indices follow the ordering shown in figure~\ref{fig:configuration}, where sensor 1 is located at the trailing edge on the upper surface, and the numbering proceeds counterclockwise around the airfoil, ending at the trailing edge on the lower surface. Each observation mode is a weighted combination of sensor outputs that, through the Kalman gain, most strongly influences the filter correction in a corresponding latent direction. At each time instant, the observation eigenmodes are ordered by descending eigenvalue magnitude, representing the most to least informative directions in the observation space. For instance, $u_1$ reflects the most informative observation direction. The weight assigned to each sensor in these modes reflects its relative contribution to that direction in the observation space.

From these plots, we observe that during the critical time interval $t \in [3.6,4.0]$, when the gust approaches and begins interacting with the airfoil on the pressure side, the leading-edge sensors 6 through 9 on the lower surface become the most responsive along $u_1$ direction. These sensors exhibit consistently high contributions across the dominant observation modes, indicating their central role in informing the estimator throughout the gust-airfoil-wake interaction ($t \in [3.6,5.0]$). While the relative contributions of sensors vary over time, the leading-edge and lower-surface sensors consistently remain the most informative along the dominant energetic mode. In contrast, the subsequent modes, $u_2$ and $u_3$, amplify the influence of the suction-side sensors, reflecting the deformation of the primary negative vortex induced by the gust.
The temporal evolution of the modes may not appear entirely smooth at first glance. However, as the gust interacts with the airfoil, the eigenvalue spectrum of the observation Gramian evolves dynamically, occasionally leading to eigenvalue crossings and a subsequent reordering of the ranked modes. For instance, the mode labeled $u_3$ at $t=3.6$ may become mode $u_2$ at the next time $t=4.0$, while the previous $u_2$ shifts to a lower rank $u_3$. Or $u_3$ at $t=4.4$ becomes $u_2$ at $t=4.8$. This ``mode reordering'' has been consistently observed across all flow cases and reflects a physical reorganization of the sensor sensitivities---that is, a temporal shift in which sensor combinations dominate the information content.

It is worth emphasizing that this observation contrasts with findings from our earlier study \citep{mousavi2025low}, in which the upstream disturbance was modeled using a Taylor vortex. A Taylor vortex typically introduces a stronger and more disruptive disturbance compared to a vortex dipole, leading to intense interaction with the primary shear layer and edge vortices on the suction side. As a result, the suction-side sensors were found to be highly sensitive and informative in that setting. In contrast, the dipole-like gust generated by Gaussian forcing in the present work interacts more locally with the pressure side, producing a more focused and asymmetric sensor response. This distinction highlights the importance of gust structure in shaping the spatial distribution of sensor informativeness and ultimately determining the directions along which the latent states can be effectively constrained.

\begin{figure*}
\centering
\includegraphics[width=0.9\textwidth]{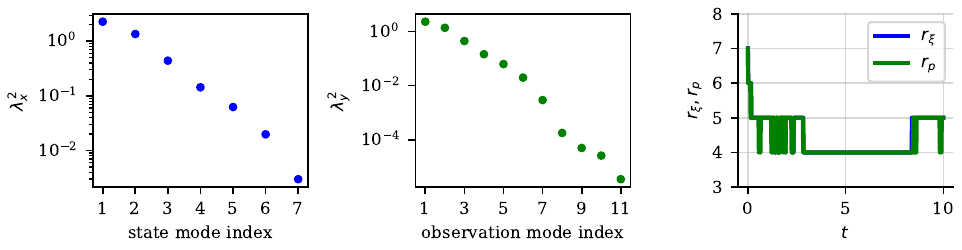}
\caption{\label{fig:AoA20_eig_rank} The left two panels display the eigenvalue spectra of the state-space and observation-space Gramians at $t=2.4$. The right plot shows the corresponding effective state rank $\stateRank$ and observation rank $\obsRank$, defined as the minimum number of eigendirections required to retain $99 \%$ of the cumulative spectral energy. The results are shown for the disturbed aerodynamic case with ($\alpha=20^\circ$, $D_y=-1.9$, $\sigma=0.1$, $y_o=0.22$, $t_o=1.5$).}
\end{figure*}

\rev{For the disturbed flow estimated in \mbox{figure~\ref{fig:estimation_AoA20}}, the spectral decomposition of the state-space and observation-space Gramians at a representative time during the gust–airfoil interaction is shown in the left two panels of \mbox{figure~\ref{fig:AoA20_eig_rank}}. The eigenvalues are arranged in descending order, corresponding to decreasing informativeness. The decay of the spectra indicates that the tail eigenmodes possess signal-to-noise ratios well below unity. These tail modes are therefore only weakly constrained by the pressure measurements, as the measurement-induced signals they generate lie within the uncertainty bounds of the sensors. Consequently, the state evolution along these weakly observed directions is governed primarily by the forecast propagation, with only negligible corrections introduced during the analysis step. Retaining $99 \%$ of the cumulative spectral energy yields the effective state rank $\stateRank$ and observation rank $\obsRank$ shown in the right plot of \mbox{figure~\ref{fig:AoA20_eig_rank}}. The state rank reflects the number of latent directions that are actively informed by the measurements and significantly corrected during assimilation, whereas the observation rank quantifies the number of independent pressure modes that effectively contribute to state updates. At early times, when the ensemble exhibits substantial deviation from the true state, corrections occur across all latent directions ($\stateRank=7$). Once the state estimate converges toward the true trajectory, the number of dominant effective modes decreases and stabilizes, oscillating between four and five as the flow evolves.}

Here, the terms ``observed'' and ``weakly observed'' are used to refer to latent directions that are strongly and weakly constrained by pressure measurements, respectively. The former correspond to modes with high observability, while the latter lie close to the approximate nullspace of the Jacobian of the observation operator, $\nabla \obs$, and therefore receive only minimal correction from the sensors.
\rev{An important conclusion is that even the weakly observed tail modes are necessary to accurately reconstruct the vorticity field. Although corrections along weak modes are small at each individual assimilation step, their cumulative effect over time improves the latent state estimate. Since the decoder maps a low-dimensional latent vector to a high-dimensional vorticity field (on the order of $2.8\times10^4$ degrees of freedom), even small latent corrections can yield noticeable improvements in the reconstructed flow.}

\begin{figure*}
\centering
\includegraphics[width=1.0\textwidth]{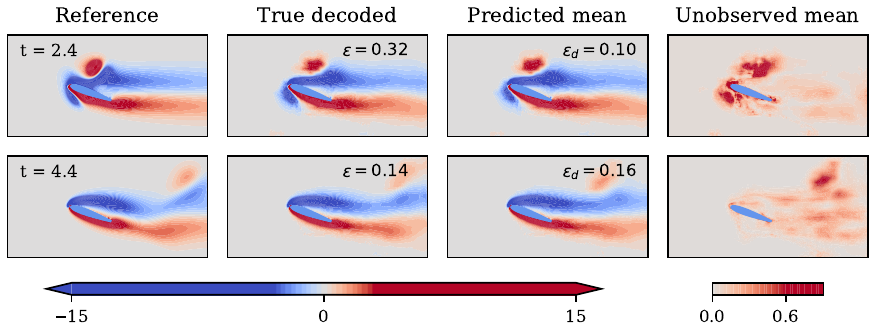}
\caption{\label{fig:unobserved_vor_AoA20} The weakly-observed vorticity plots for disturbed flow at condition ($\alpha=20^\circ$, $D_y=-1.9$, $\sigma=0.1$, $y_o=0.22$, $t_o=1.5$).}
\end{figure*}

For the case $\alpha=20^\circ$ under the same disturbed condition shown in figure~\ref{fig:estimation_AoA20}, we further examine these weakly observed directions (corresponding to the smallest eigenvalues in figure \ref{fig:AoA20_eig_rank}) by explicitly constructing perturbations in the approximate nullspace of \rev{$\obsGram$ (which corresponds to the nullspace of $\nabla \obs$ in the deterministic limit)}. The resulting vorticity reconstructions, shown in figure~\ref{fig:unobserved_vor_AoA20}, demonstrate that perturbations along these latent directions can produce predicted pressure measurements that remain within sensor uncertainty bounds, but still correspond to notable variation in the flow fields.
This provides direct evidence of the many-to-one mapping between latent representations and pressure signatures. The induced vorticity differences are concentrated primarily in regions of limited sensor sensitivity, including the gust core, the leading-edge vicinity, the suction-side shear layer, and the wake. It confirms that weakly observed latent modes can contribute significantly to overall vorticity error.
These results underscore a fundamental limitation of pressure-only sensing: in the absence of additional measurement modalities or prior regularization, multiple latent states may remain indistinguishable from the perspective of the sensors. This ambiguity becomes more pronounced after the gust convects downstream, when pressure sensitivity to residual flow variations decreases. Consequently, multiple plausible latent configurations may be indistinguishable from the available measurements during such regimes, and it increasingly falls on the forecast operator to propagate the latent state.

To mitigate this ambiguity, we explored incorporating a maximum-observability constraint during autoencoder training (section~\ref{sec:data-compression}) to enforce a more uniquely pressure-consistent latent representation. While this approach improved latent state identifiability, it rendered the vorticity decoder highly sensitive to small latent perturbations, such that minor estimation errors produced non-physical artifacts and degraded reconstruction robustness. This tradeoff highlights the inherent tension between enhancing measurement-space observability and maintaining stability and robustness in high-dimensional flow reconstruction within compressed latent-variable frameworks.

\subsection{Sensor failure study}
As shown in the preceding analysis, leading-edge sensors (sensors 6 and 7) play a dominant role in informing the data assimilation process. A natural set of questions arises: What happens if a highly informative sensor fails or is removed? How significantly would this affect the accuracy of the state estimation? Would the estimator become unreliable simply due to the loss of a single, influential sensor?
To address these questions, we investigate the effect of sensor dropout on flow estimation. A sensor can be effectively ``dropped" by assigning it a much larger measurement noise variance, thereby diminishing its contribution in the assimilation update. Specifically, we simulate the failure of sensor 7, a leading-edge sensor identified as highly informative along $u_1$ in figure~\ref{fig:obs_modes_AoA60}, by increasing its variance by a factor of 1000 relative to all other sensors. Data assimilation is then performed for the same flow condition presented in figure~\ref{fig:obs_modes_AoA60}. For comparison, we also drop sensor 2, located near the trailing edge (TE) on the suction side---an area associated with minimal sensor response---in a separate setting and assess its impact relative to a baseline case where all sensors have equal uncertainty (see figures~\ref{fig:estimation_AoA60} and \ref{fig:obs_modes_AoA60}). 

\begin{figure}[tb]
\centering
\includegraphics[width=0.95\textwidth]{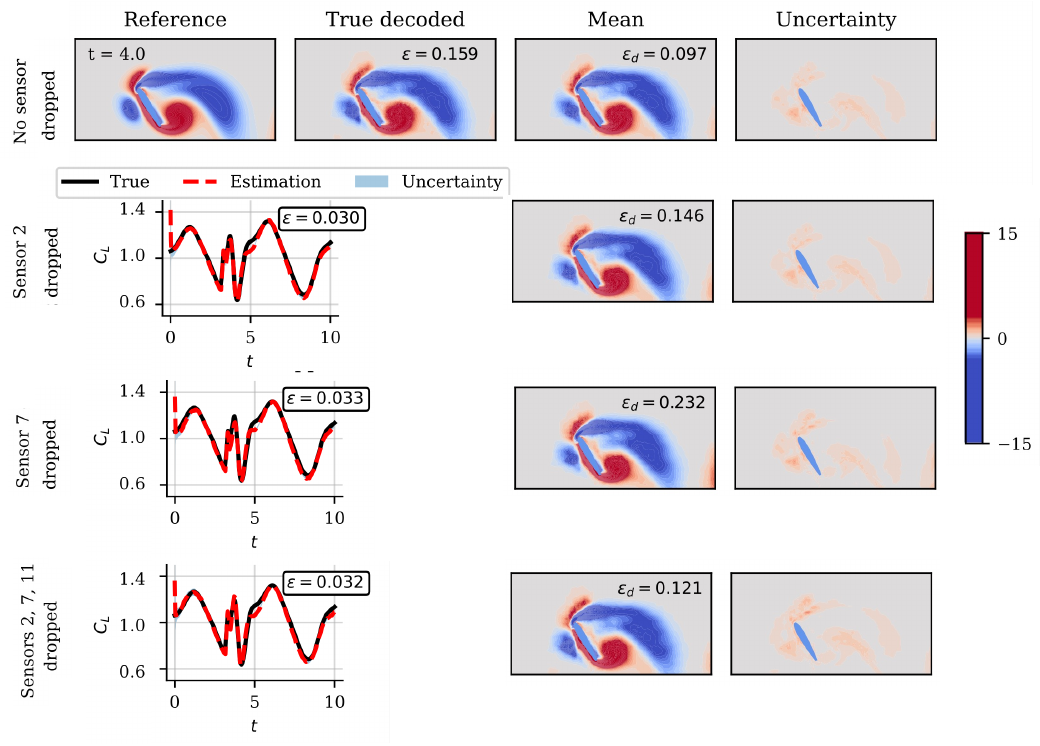}
\caption{\label{fig:sensor_dropout} Effect of individual sensor dropout on the reconstructed lift and vorticity fields for the case ($\alpha=60^\circ$, $D_y=-0.71$, $\sigma=0.12$, $y_o=-0.19$, $t_o=3.3$). }
\end{figure}

\begin{figure}[tb]
\centering
\includegraphics[width=0.95\textwidth]{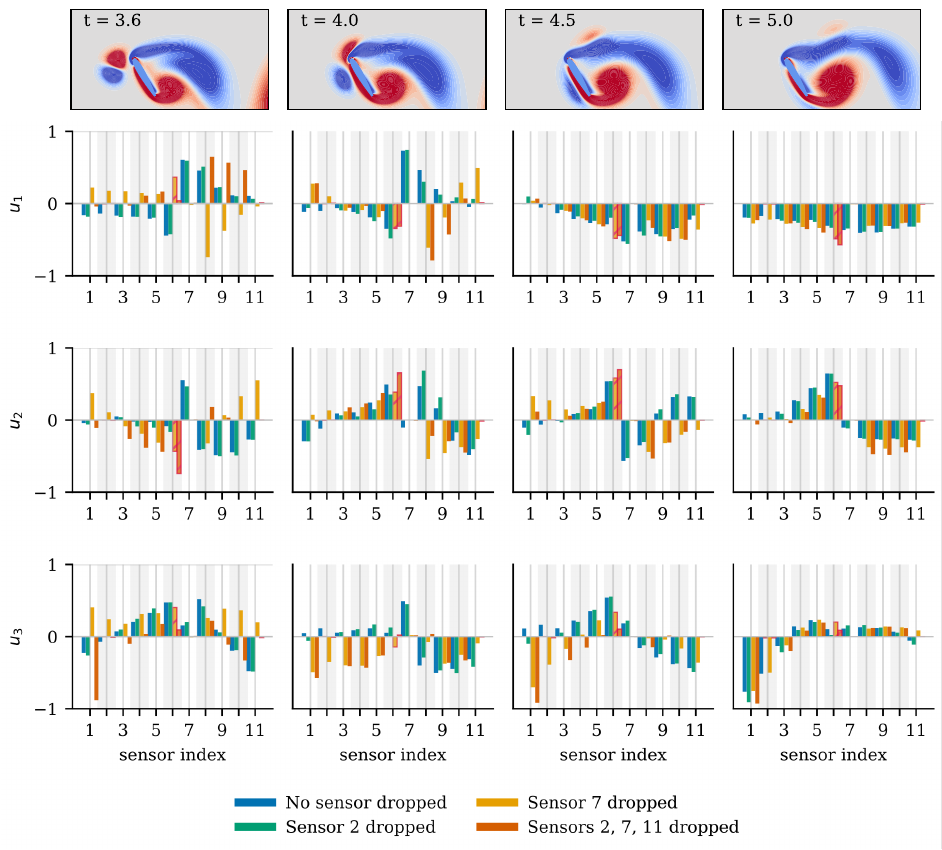}
\caption{\label{fig:sensor_dropout_obs_modes} Effect of individual sensor dropout on the observation modes consistent with figure~\ref{fig:sensor_dropout} at four representative time instances during the gust-airfoil interaction. The condition is ($\alpha=60^\circ$, $D_y=-0.71$, $\sigma=0.12$, $y_o=-0.19$, $t_o=3.3$).}
\end{figure}

The results shown in figure~\ref{fig:sensor_dropout} reveal that dropping either sensor leads to only a modest increase in the average lift prediction error (this error is $\varepsilon=0.03$ for the case of no sensor being dropped), indicating that the estimator remains robust in capturing global aerodynamic loads. However, the impact on the reconstructed vorticity field is more pronounced, particularly when sensor 7---located near the leading edge (LE) on the pressure side---is removed. In this case, even with greater reconstruction error, the estimated vorticity field still captures the large-scale structures in the domain.
In contrast, dropping sensor 2---located near the TE on the suction side---has a negligible effect on both the reconstructed flow field and the associated uncertainty bounds. This suggests that sensor 2 contributes minimally to the information content of the observation operator in this particular flow regime and sensor configuration. The removal of sensor 7, which had a strong contribution to the leading observation modes (as seen in figure~\ref{fig:obs_modes_AoA60}), significantly alters the structure of the dominant eigenvectors of the observation Gramian $\obsGram$, evident in figure~\ref{fig:sensor_dropout_obs_modes}. Specifically, these observation eigenmodes reconfigure by suppressing the weight of sensor 7 to nearly zero and compensating by reallocating higher weights to its neighboring sensors. This redistribution allows the estimator to partially recover the lost observability; however, the reconfigured observation space may not fully span the same latent directions as before.
\rev{It should be noted that although sensor 7 is highly informative in the leading observation eigenmode, other sensors carry comparable—sometimes even greater—information in the subsequent leading modes, as shown in \mbox{figure~\ref{fig:obs_modes_AoA60}}, and thus can compensate the loss of sensor 7.}

To further assess the estimator’s robustness, we examine the scenario in which multiple sensors, here sensors 2, 7, and 11, fail simultaneously. The corresponding results are shown in figures~\ref{fig:sensor_dropout} and \ref{fig:sensor_dropout_obs_modes}. To recall, sensor 7 originally provided the most informative measurements in the most energetic eigendirection $u_1$, so the overall reconstruction quality is comparable to the case where only sensor 7 is removed, evident in the dominant observation modes illustrated in figure~\ref{fig:sensor_dropout_obs_modes}. Interestingly, the vorticity reconstruction is slightly improved in the three-sensor failure case, which may be attributed to sensors 2 and 11 being highly correlated with other active sensors. Their removal can, at certain instants, reduce redundancy in the observation space and improve numerical conditioning, thereby marginally enhancing the filter’s performance in vorticity predictions. Interestingly, despite the simultaneous failure of 3 out of 11 sparse sensors (representing a $27\%$ outage), the estimator maintains robust performance.

Overall, by examining how dominant observation modes evolve under sensor dropout, one can identify the relative importance of each sensor, assess redundancy in the sensing architecture, and quantify the robustness of the estimation framework to partial sensor failure. The findings confirm that the current approach maintains resilience to sensor loss, with minimal degradation in flow estimation accuracy. Even in the case of dropping a highly informative sensor---such as sensor 7 near the LE---the estimator adapts by reallocating observational weight to neighboring sensors and continues to perform well in predicting global quantities like lift and capturing the dominant large-scale vortical structures in the flow. This highlights the robustness of the learned filtering framework and its potential for deployment in scenarios with limited or partially compromised sensing.

\subsection{Sequential estimation for an extrapolated case}
\begin{figure}[tb]
\centering
\includegraphics[width=1.0\textwidth]{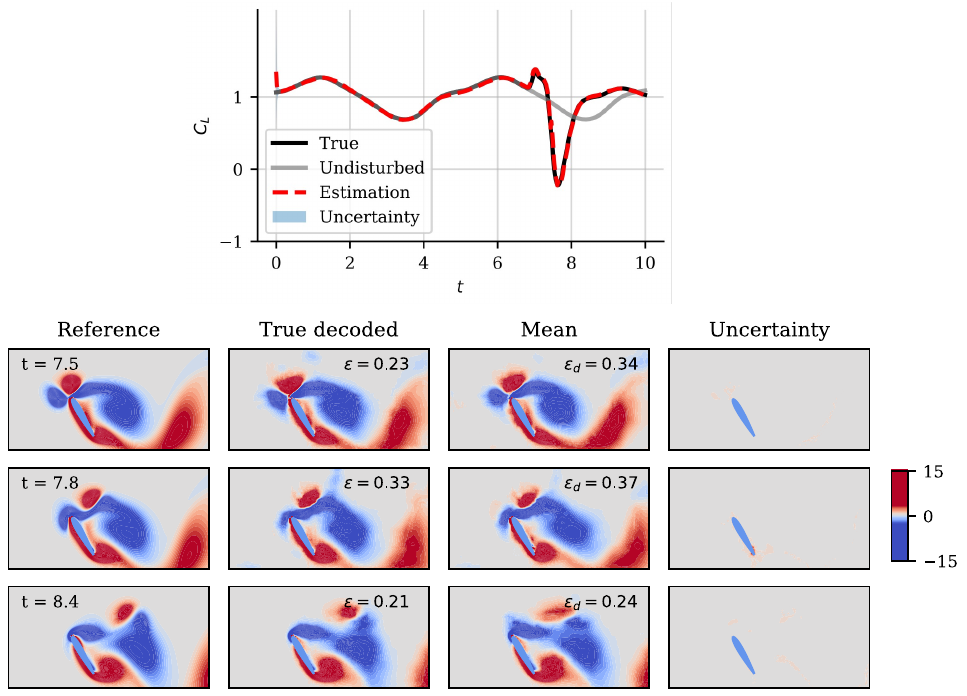}
\caption{\label{fig:extrapolation_AoA60} Estimated lift and vorticity field for disturbed aerodynamics at the extrapolation condition of ($\alpha=60^\circ$. $D_y=-1.93$, $\sigma=0.1$, $y_o=0.22$, $t_o=7.0$). The predicted ensemble mean along with the $95\%$ confidence interval is shown for the lift coefficient and the vorticity field.}
\end{figure}

All training and test aerodynamic cases discussed thus far involved disturbances introduced during the first vortex shedding cycle. To evaluate the robustness and generalizability of the estimator, it is critical to test its performance under extrapolative conditions---specifically when a disturbance occurs at an unseen phase of the shedding cycle. For this purpose, we construct a test case where a dipole-like gust is introduced at $t_o=7$, well into the second shedding cycle, at the maximum AoA considered in this study, $\alpha = 60^\circ$. We refer to this as the extrapolation case. The results are presented in figure~\ref{fig:extrapolation_AoA60}, showing the predicted lift coefficient and vorticity fields at three representative time instances during the gust–airfoil interaction. Before the gust enters the domain, the predicted lift trajectory matches the undisturbed periodic behavior with excellent accuracy, reflecting the estimator's ability to maintain nominal performance under regular flow conditions. When the gust is introduced at $t_o=7$, the pressure sensors immediately detect the upstream disturbance, resulting in a clear deviation from the undisturbed lift trajectory. This reflects the pressure field’s sensitivity to flow perturbations, even before their direct interaction with the airfoil around $t=7.4$. Importantly, the forecast operator is incapable of anticipating the effect of an incoming gust at $t_o=7$. As such, the forecasted latent states $\lat (t_o+\Delta t|t_o)$ remain confined to the undisturbed latent manifold. The deviation from undisturbed behavior is therefore initiated entirely by the correction introduced in the analysis step of filtering, wherein the discrepancy between observed and predicted pressures triggers an adjustment to the predicted states, resulting in the updated posterior $\lat (t_o+\Delta t|t_o+\Delta t)$. 

The reconstructed vorticity field reflects this behavior. The posterior mean closely captures the dominant shear layers and large-scale vortices. The temporal evolution of the flow is accurately tracked, including the unsteady shedding patterns and the influence of the gust. Notably, the estimated uncertainty fields are so small that they are visually not detectable. This extrapolation test confirms that the filtering framework generalizes well beyond its training window, relying on pressure-based corrections to capture flow disturbances introduced at unseen phases. The pressure sensors effectively detect and drive state corrections quickly, while the uncertainty quantification provides meaningful confidence estimates. This underscores the estimator's reliability and adaptability in operational settings where disturbances may arise unpredictably. 

One striking and consistent phenomenon observed across all estimated cases in Figures~\ref{fig:estimation_AoA20}, \ref{fig:estimation_AoA60}, and \ref{fig:extrapolation_AoA60} (as mentioned earlier) is the transient lift spike caused by the interaction between the dipole gust and the boundary layers. Notably, the largest estimation error and the highest uncertainty magnitude consistently occur during the time of maximum lift deviation from the undisturbed trajectory---precisely when the dipole is in closest interaction with the primary edge vortex. This phase corresponds to the most dynamically complex portion of the flow evolution, characterized by intense vortex interactions, rapid structural changes in the shear layer, and strong nonlinearity in the pressure response. The estimator reflects this complexity by exhibiting both increased prediction error and broader uncertainty bounds, indicating reduced observability and greater difficulty in capturing the rapidly changing flow state.

Across all test cases, the estimator maintains accurate lift tracking and resolves the dominant vortical structures from sparse pressure measurements. By embedding NN surrogates within an ensemble filtering architecture, the method not only achieves computational efficiency but also adapts naturally to degraded sensing and extrapolative flow conditions. The result is an estimator that remains robust under severe disturbances, provides interpretable uncertainty quantification, and scales efficiently for online deployment. Such capabilities establish a foundation for practical integration of flow state monitoring with feedback control and decision-making in unsteady aerodynamic environments.

\section{Conclusion}\label{sec:conclusion}
This work proposed a data assimilation framework in a reduced space for fast flow and load estimation in randomly disturbed aerodynamic environments. By combining nonlinear convolutional autoencoders for compact state representation, neural-network-based surrogate models for dynamics and observations, and ensemble Kalman filtering (EnKF), the proposed approach enables accurate, uncertainty-aware estimation of unsteady flow fields from sparse, noisy surface pressure measurements. \rev{It was discussed that the learned forecast operator does not capture the transition from undisturbed to disturbed trajectories without measurement correction. In the absence of assimilation, the forecast predicts that the latent state remains on the nominal baseline trajectory. The measurement update step introduces the deviation from this trajectory once the pressure signature of the disturbance appears, enabling the estimator to track the disturbed evolution.} The estimator was evaluated across a wide range of gust-disturbed aerodynamic scenarios, including both interpolative cases---where disturbances occurred during the first vortex shedding cycle, consistent with training---and extrapolative cases---where gusts were introduced at arbitrary instants across the shedding cycle. In all test scenarios, the predicted lift closely followed the true unsteady dynamics, and the reconstructed vorticity fields accurately captured the dominant vortical structures, especially in regions adjacent to sensor locations. The estimator demonstrated strong generalization capabilities, reliably tracking complex flow behavior under conditions not encountered during training.

To quantify the limits of observability, we performed eigenvalue decompositions of the state-space and observation-space Gramians. Across all cases, the system exhibits intrinsically weakly observed directions under surface pressure sensing. In the original vorticity space, these directions correspond primarily to the gust core and portions of the downstream wake. These regions were shown to lie approximately in the nullspace of the Jacobian of the observation operator, indicating that perturbations along these directions produce negligible pressure signatures. We further analyzed sensor informativeness over time by computing the dominant modes of the observation Gramian, which revealed the transient sensitivity of each pressure sensor to the evolving latent state. Leading-edge sensors were consistently the most informative in the first few energetic modes, especially during gust–vortex interactions, while suction-side sensors near the trailing edge contributed less throughout the flow evolution. Sensor informativeness during gust encounters strongly depends on the gust structure. For example, in our previous work \citep{mousavi2025low}, a Taylor gust produced a more disruptive interaction, leading to different information content across sensors compared to the less disruptive and more compact Gaussian gust considered in this study. It is important to note that the EnKF employed in this study assumes all sources of uncertainty to be Gaussian or well-approximated as such. Nonetheless, for strongly nonlinear and non-Gaussian systems, this framework can be extended using the gradient-based Gramian estimation approach proposed by \citet{baptista2022gradient}.

To probe the estimator’s robustness, we studied the impact of sensor dropout by artificially increasing the uncertainty of key sensors, effectively removing them from the assimilation process. Remarkably, even when the most informative leading-edge sensor was dropped, the estimator maintained high accuracy. This resilience stems from the mode redistribution of the system: when a dominant sensor is lost, the filter implicitly increases the Kalman gain weights on neighboring sensors, allowing them to compensate and reconstruct the dominant flow features. As a result, the degradation in lift and vorticity estimation was minimal, confirming that the estimator leverages spatial redundancy and shared observability among nearby sensors.

The gust–vortex interactions corresponded to the most dynamically complex phases of the flow evolution and were consistently associated with the largest estimation errors and highest posterior uncertainty. The uncertainty fields correctly identified regions of physically-limited observability---such as the gust core and wake shear layers---indicating that the estimator not only reconstructs the dominant flow features but also provides reliable confidence bounds.
Overall, the results demonstrate that the proposed framework achieves accurate and efficient flow state estimation under strong arbitrary disturbances, with robustness to sensor dropout and generalization to unseen gust realizations. By exploiting sequential filtering, this method provides a scalable and reliable foundation for flow monitoring, feedback control, and decision-making in aerodynamic systems where traditional solvers are computationally prohibitive.

A critical component in bridging sparse pressure observations with high-dimensional flow state estimation is the construction of a low-dimensional latent representation. However, as discussed in this study, the resulting latent space lacks its own physical interpretability \rev{without the use of the encoder to lift it back to the full flow space}, making it challenging to relate individual latent components to specific flow structures or phenomena. Future work may focus on enhancing interpretability---either through targeted regularization strategies or by modifying the autoencoder architecture---to promote more physically meaningful latent variables and improve transparency in the estimation process. It should be emphasized that this study explored an extreme scenario in which the forecast operator lacks long-time accuracy, demonstrating that even under such conditions, surface pressure measurements contain sufficient flow information to effectively detect arbitrary gusts and correct the predicted states through filtering. However, the limitations imposed by weakly-observed latent directions---those that cannot be recovered from pressure data alone---can be mitigated by either learning a more accurate forecast operator, which provides a stronger prior for the filtering process, or by augmenting the measurement vector with other sensor data. This will be addressed in our future work. Finally, this framework is broadly applicable to systems where a reduced-order state representation can be constructed.

\appendix
\section{Update step of the ensemble Kalman filter} \label{sec:apkalman}
\renewcommand{\theequation}{\thesection.\arabic{equation}}
\setcounter{equation}{0}
For the update formula defined in Eq.~\eqref{eq:kalman_update}, the Kalman gain from the standard (linear) Kalman filter is
\begin{align} \label{apeq:gain}
    \gain &= \stateObsCov \, \obsCrossCov^{-1} \\
    &= \priorCov \hlin^{\top} \left( \hlin \priorCov \hlin^{\top} + \obsCov \right)^{-1},
\end{align}
where $\stateObsCov$ and $\obsCrossCov$ represent the state-observation cross-covariance and observation covariance matrices, respectively, $\priorCov$ is the prior covariance, $\hlin$ is the linear observation operator (or the tangent linear of a nonlinear observation operator, such as $\obs$ in the present study), and $\obsCov$ is the measurement-noise covariance defined in Section~\ref{sec:filtering}. Here, we aim to derive the ensemble approximation of these covariances.

We use the subscripts $f$ and $a$ to denote forecast (prior) and analysis (posterior) quantities, respectively. Given an ensemble of state $\{\lat^i\}_{i=1}^M$ with $\lat^i\in\mathbb{R}^n$, define the ensemble matrix $\Lat = [\lat^1,\dots,\lat^M]\in\mathbb{R}^{n\times M}$ and the ensemble mean
\begin{equation}\label{apeq:mean}
    \bar{\lat} = \frac{1}{M} \sum_{i=1}^M \lat^i.
\end{equation}
Let $\Lat^\prime$ be the normalized anomaly matrix with columns
\begin{equation} \label{apeq:anomaly}
    [\Lat^\prime]^i = \frac{1}{\sqrt{M-1}} ( \lat^i - \bar{\lat}).
\end{equation}
These definitions apply to both the forecast and analysis ensembles, yielding $\bar{\lat}_f$, $\Lat_f^\prime$, $\bar{\lat}_a$, and $\Lat_a^\prime$.

With the perturbed observation vector as defined in Eq.~\eqref{eq:observation}, where $\obsnoise^i \sim \mathcal{N}(\pmb{0}, \obsCov)$, let us define the innovation anomaly matrix with $i$th column given by
\begin{equation}
    [\Pres^\prime_f]^i = \frac{\obs(\lat_f^i) + \obsnoise^i - \obs(\bar{\lat}_f) - \bar{\obsnoise}}{\sqrt{M-1}},
\end{equation}
with $\bar{\obsnoise}$ the sample mean of $\{\obsnoise^i \}$. It should be noted that this mean is itself a random number whose expected value is 0. 

To connect the Kalman filter gain \eqref{apeq:gain} to its ensemble form, we follow \citet{asch2016data} and employ the standard linearization:
\begin{equation}
    \hlin (\lat_f^i - \bar{\lat}_f) \simeq \obs(\lat_f^i) - \bar{\pres}_f.
\end{equation}
Using this approximation, it is then easy to show that \citep{asch2016data}:
\begin{equation} \label{apeq:stateobscov}
    \priorCov \hlin^{\top} \simeq \Lat^{\prime}_f \Pres^{\prime \top}_f = \text{Cov}(\Lat_f, \Pres_f),
\end{equation}
and 
\begin{equation} \label{apeq:obsobscov}
    \left( \hlin \priorCov \hlin^{\top} + \obsCov \right) \simeq \Pres^{\prime}_f \Pres^{\prime \top}_f = \text{Cov}(\Pres_f, \Pres_f).
\end{equation}

\section*{Acknowledgments}

The authors gratefully acknowledge the financial support provided by the National Science Foundation under award numbers 2247005 and 2247006.

\bibliographystyle{unsrtnat}
\bibliography{refs}

\end{document}